\documentclass{aa501}     
\usepackage{graphics}
 
\def\cc{\,{\rm cm^{-3}}}
\def\cm2{\,{\rm cm^{-2}}}
\def\kms{\,{\rm {km\,s^{-1}}}}
\def\kkms{\,{\rm {K\,km s^{-1}}}}

\def\co{\,{\rm ^{12}CO}}
\def\13co{\,{\rm ^{13}CO}}
\def\h2{\,{\rm H_{2}}}
\def\aua{{\rm A\&A} }

\def\apj{{\rm ApJ} }
\def\aj{{\rm AJ} }
\def\apjs{{\rm ApJS} }
\def\apjl{{\rm ApJL} }

\def\pasj{{\rm PASJ} }
\begin{document}
 
\title{CI and CO in the spiral galaxies NGC~6946 and M~83}
 
   \subtitle{}
 
\author{F.P. Israel
          \inst{1}
           and F. Baas
          \inst{1,2}
           }
 
   \offprints{F.P. Israel}
 
   \institute{Sterrewacht Leiden, P.O. Box 9513, 2300 RA Leiden,
             The Netherlands
   \and       Joint Astronomy Centre, 660 N. A'ohoku Pl., Hilo,
             Hawaii, 96720, USA}
 
\date{Received ????; accepted ????}
 
\abstract
{We present $J$=2--1, $J$=3--2, $J$=4--3 $^{12}$CO and 492 GHz [CI] maps as
well as $J$=2--1 and $J$=3--2 $^{13}$CO measurements of the late type spiral 
galaxies NGC~6946 and M~83 (NGC~5236). Both galaxies contain a pronounced
molecular gas concentration in rapid solid-body rotation within a few 
hundred parsec from their nucleus. NGC~6946 and M~83 have nearly identical 
relative intensities in the $^{12}$CO, $^{13}$CO and [CI] transitions, but 
very different [CII] intensities, illustrating the need for caution in 
interpreting CO observations alone. The slow decrease of velocity-integrated 
$^{12}$CO intensities with increasing rotational level implies the presence 
of significant amounts of warm and dense molecular gas in both galaxy centers. 
Detailed modelling of the observed line ratios indicates that the molecular 
medium in both galaxies consists of at least two separate components. These 
are a warm and dense component ($T_{\rm kin}$ = 30 -- 60 K, $n(\h2) = 
3000 - 10 000 \cc$) and a much more tenuous hot component ($T_{\rm kin}$ = 
100 -- 150 K, $n(\h2) \leq 1000 \cc$). Total atomic carbon column 
densities exceed CO column densities by a factor of about 1.5 in NGC~6946 
and about 4 in M~83. Unlike NGC~6946, M~83 contains a significant amount of 
molecular hydrogen associated with ionized carbon rather than CO. The centers 
of NGC~6946 and M~83 contain nearly identical total (atomic and molecular) 
gas masses of about 3 $\times 10^{7}$ $M_{\odot}$. Despite their prominence, 
the central gas concentrations in these galaxies represent only a few per 
cent of the stellar mass in the same volume. The peak face-on gas mass 
density is much higher in M~83 (120 M$_{\odot}$\, pc$^{-2}$) than in 
NGC~6946 (45 M$_{\odot}$\, pc$^{-2}$). The more intense starburst in M~83 
is associated with a more compact and somewhat hotter PDR zone than the 
milder starburst in NGC~6946.
}

\maketitle
 
\section{Introduction}

\begin{table}[h]
\caption[]{Galaxy parameters}
\begin{flushleft}
\begin{tabular}{lll}
\hline
\noalign{\smallskip}
			    & M~83				  & NGC~6946 \\
\noalign{\smallskip}
\hline
\noalign{\smallskip}
Type$^{a}$     	 	    & SBc				  & Scd \\
Optical Centre:		    & 					  & \\
R.A. (1950)$^{b}$ 	    & 13$^{h}$34$^{m}$11.6$^{s}$  	  & 20$^{h}$33$^{m}$48.8$^{s}$ \\
Decl.(1950)$^{b}$           & -29$^{\circ}$36$'$42$''$ 	          & +59$^{\circ}$58$'$50$''$ \\
Radio Centre :		    & \\
R.A. (1950)$^{c}$ 	    & 13$^{h}$34$^{m}$11.1$^{s}$	  & 20$^{h}$33$^{m}$49.1$^{s}$ \\
Decl.(1950)$^{c}$           & -29$^{\circ}$36$'$34.9$''$	  & +59$^{\circ}$58$'$49$''$ \\
$V_{\rm LSR}^{d}$    	    & 510 $\kms$ 			  & 55 $\kms$ \\
Distance $D^{e}$            & 3.5 Mpc 				  & 5.5 Mpc\\
Inclination $i^{d}$ 	    & 24$^{\circ}$ 			  & 38$^{\circ}$ \\
Position angle $P^{d}$      & 45$^{\circ}$ 			  & 60$^{\circ}$ \\
Luminosity $L_{\rm B}^{e}$  & 1.2 $\times 10^{10}$ L$_{\rm B\odot}$ & 3 $\times 10^{10}$ L$_{\rm B\odot}$ \\
Scale           	    & 59 $''$/kpc 			  & 38 $''$/kpc \\
\noalign{\smallskip}
\hline
\end{tabular}
\end{flushleft}
Notes to Table 1:\\
$^{a}$ RSA (Sandage $\&$ Tammann 1987);
$^{b}$ Dressel $\&$ Condon (1976);Rumstay $\&$ Kaufman 1983;
$^{c}$ Turner $\&$ Ho 1994; van der Kruit et al. (1977);
$^{d}$ Tilanus $\&$ Allen 1993; Handa et al. (1990); Carignan et al. (1990);
$^{e}$ Banks et al. 1999; Tully (1988);
\end{table}

Molecular gas is a major constituent of the interstellar medium in galaxies
and the dominating component in regions of star formation and the inner disks 
of spiral galaxies. Within the inner kiloparsec, many spiral galaxies also
exhibit a strong concentration of molecular gas towards their nucleus.
It is generally thought that such 
concentrations are the result of angular momentum losses caused by e.g.
encounters or mergers with other galaxies, or by bar-like potentials
in the central part of the galaxy. However, in some cases, such as the Sb 
spiral galaxies M~31 and NGC~7331, most or all of the central gas may 
have originated from mass loss by evolved stars in the bulge (cf.
Israel $\&$ Baas 1999). In order to determine the physical condition
of molecular gas in the centers of galaxies, and its amount, we have
conducted a programme to observe a number of nearby galaxies in various
CO transitions, as well as the 492 GHz $^{3}$P$_{1}$--$^{3}$P$_{0}$ CI 
transition. Results for the Sc galaxy NGC~253 (Israel, White $\&$ Baas 1995)
and the Sb galaxy NGC~7331 (Israel $\&$ Baas 1999) have already been 
published, as well as preliminary results on the Sc galaxy NGC~3628 (Israel,
Baas $\&$ Maloney 1990). In this paper, we present the results for the Sc 
galaxies NGC~6946 and M~83. Basic properties of these galaxies are summarized
in Table 1.

Although a member of the NGC~6643 group, NGC~6946 (Arp 29) is relatively 
isolated. Its distance is variously estimated between 3 Mpc
(Ables 1971) and 10 Mpc (Rogstad $\&$ Shostak 1972; Sandage $\&$ Tammann
1974); here we adopt $D$ = 5.5 Mpc (Tully 1988; McCall 1982). It has been
relatively well-studied in the lower CO transitions. In fact, it was one of
the first galaxies mapped in $J$=1--0 CO at 65$''$ resolution (Morris $\&$ 
Lo 1978; Rickard $\&$ Palmer 1981. Higher-resolution maps at 17--23$''$
were published by Sofue et al. (1988) and Weliachew et al. (1988). At a 
similar resolution, disk spiral arm regions were observed in $J$=1--0 and 
$J$=2--1 CO by Casoli et al. (1990). The central region was also observed
in the $J$=3--2 and $J$=4--3 CO transitions, again at similar resolutions
(Wall et al. 1993; Mauersberger et al. 1999; Nieten et al. 1999). Using 
ISO, Valentijn et al. (1996) obtained a direct detection of 
warm $\h2$ towards the center of NGC~6946. Early high-resolution (about 
6.5$''$) $J$=1--0 maps were obtained by Ball et al. (1985) and Ishizuki 
et al. (1990). Very good maps with a resolution of 3--4$''$ can be found 
in Regan $\&$ Vogel (1995) and Sakamoto et al. (1999). These maps show an
elongated concentration of CO in the center, extending to the northwest
with a position angle changing from 315$^{\circ}$ close to the nucleus
to 0$^{\circ}$ at 15$''$ from the nucleus. Similarly high-resolution maps 
of nuclear HCN emission by Helfer $\&$ Blitz (1997) show only a compact source.

\begin{figure*}[]
\unitlength1cm
\begin{minipage}[b]{3.94cm}
\resizebox{4.2cm}{!}{\includegraphics*{n6946_co21.ps}}
\end{minipage}
\hfill
\begin{minipage}[t]{3.94cm}
\resizebox{4.2cm}{!}{\includegraphics*{n6946_co32.ps}}
\end{minipage}
\hfill
\begin{minipage}[t]{3.94cm}
\resizebox{4.2cm}{!}{\includegraphics*{n6946_co43.ps}}
\end{minipage}
\hfill
\begin{minipage}[t]{3.94cm}
\resizebox{4.2cm}{!}{\includegraphics*{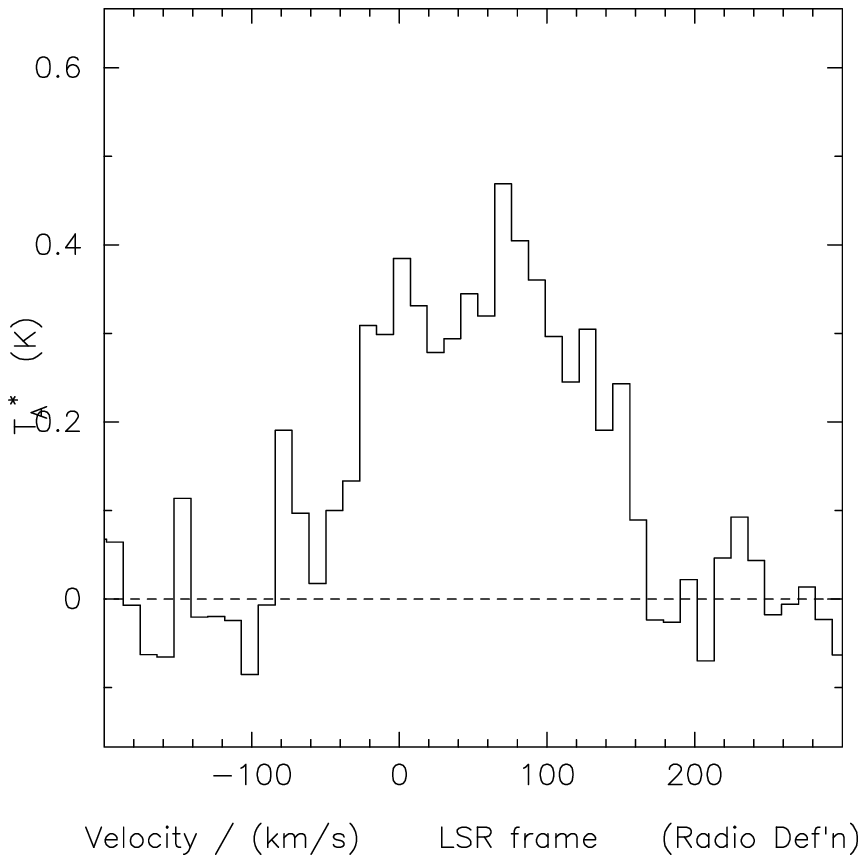}}
\end{minipage}
\begin{minipage}[b]{3.94cm}
\resizebox{4.2cm}{!}{\includegraphics*{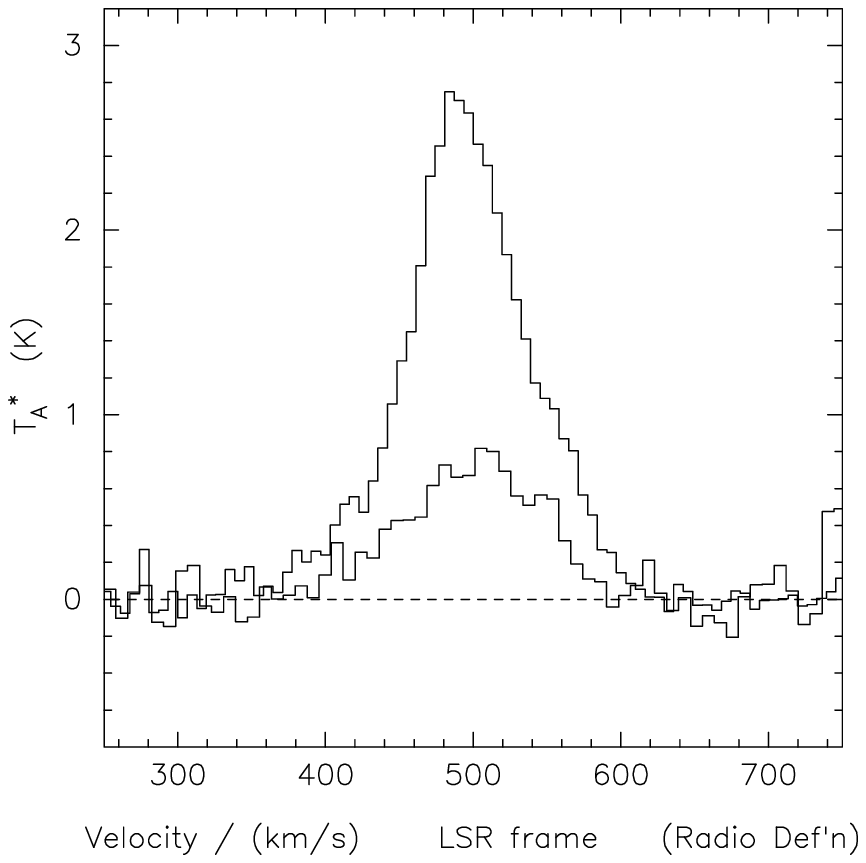}}
\end{minipage}
\hfill
\begin{minipage}[t]{3.94cm}
\resizebox{4.2cm}{!}{\includegraphics*{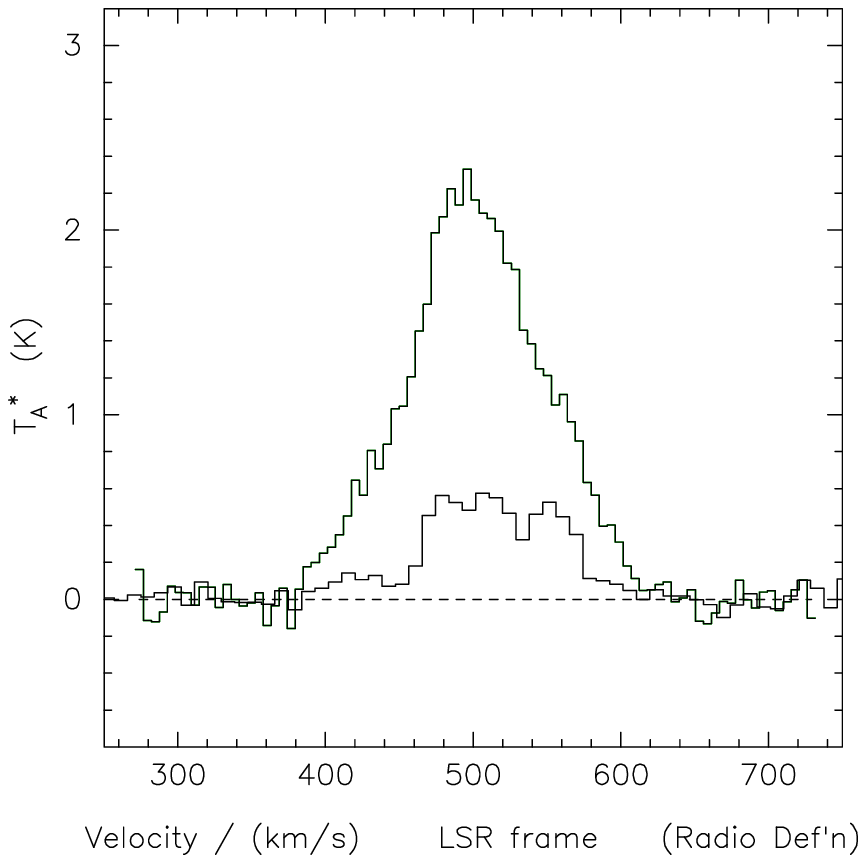}}
\end{minipage}
\hfill
\begin{minipage}[t]{3.94cm}
\resizebox{4.2cm}{!}{\includegraphics*{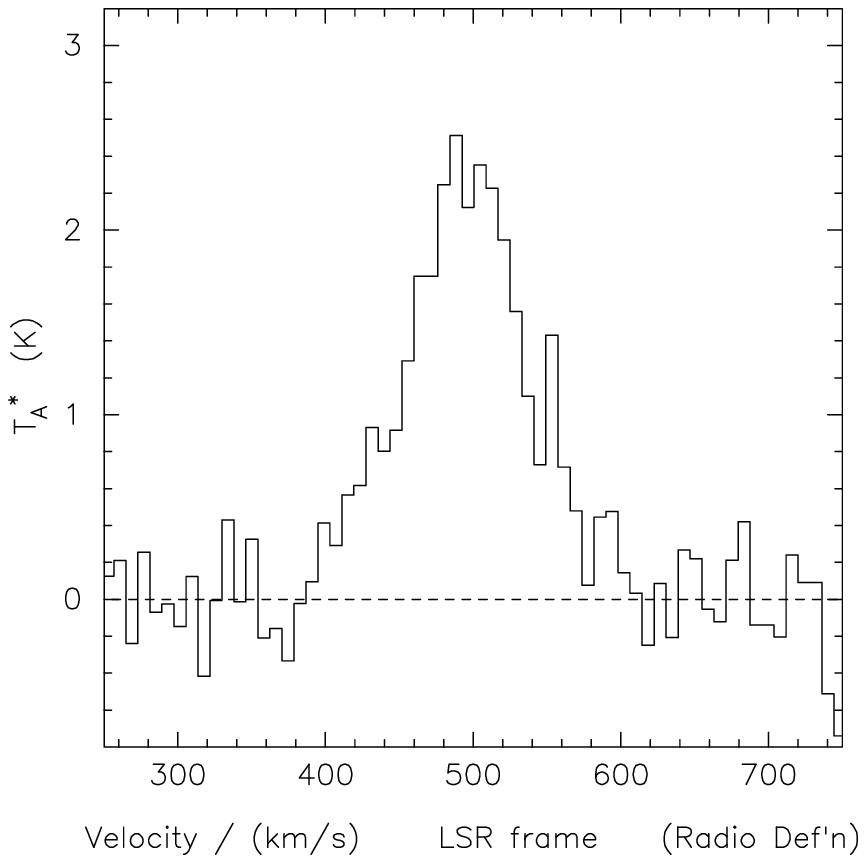}}
\end{minipage}
\hfill
\begin{minipage}[t]{3.94cm}
\resizebox{4.2cm}{!}{\includegraphics*{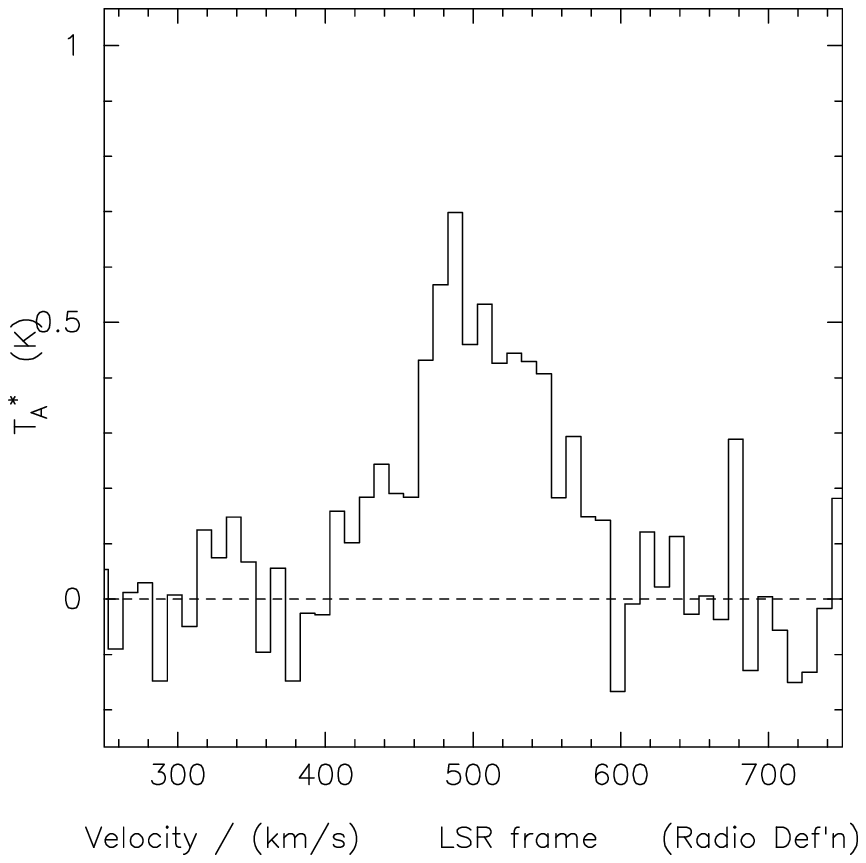}}
\end{minipage}
\caption[]
{Full resolution emission Spectra observed towards the centers of 
NGC~6946 and M~83.
Top row: NGC~6946; bottom row: M~83. Columns from left to right:
$J$=2--1 CO, $J$=3--2 CO, $J$=4--3 CO, [CI]. Vertical scale is actually 
in $T_{\rm mb}$. Whenever available, $\13co$ profiles are shown
as the lower of the two profiles in the appropriate $\co$ box, but with 
brightness temperatures multiplied by three, i.e. on the same temperature 
scale as [CI].}
\end{figure*}

\begin{table*}[]
\caption[]{Observations Log}
\begin{flushleft}
\begin{tabular}{llcccccccccc}
\hline
\noalign{\smallskip}
Transition & Object & Date    	& Freq	& $T_{\rm sys}$ & Beam 	& $\eta _{\rm mb}$ & t(int) & \multicolumn{4}{c}{Map Parameters} \\
& & & & & Size & & & Points & Size & Spacing & P.A. \\
	   &	    & (MM/YY) 	& (GHz)	& (K)	  & ($\arcsec$) & 	       	   & (sec) & & ($\arcsec$) & ($\arcsec$) & ($^{\circ}$)	\\
\noalign{\smallskip}
\hline
\noalign{\smallskip}
$^{12}$CO $J$=2--1   & NGC~6946	& 02-06/89 & 230  & 1100  	& 21	& 0.63  & 600  & 36 & 60$\times$60  & 10 & 0 \\
	   & M~83   	& 02-89    &      & 1295       	&       & 0.63  & 600  & 49 & 70$\times$120 & 10 & 45 \\
$^{12}$CO $J$=3--2   & NGC~6946	& 12/93    & 345  & 1270  	& 14	& 0.53  & 400 
& 40 & 54$\times$54  &  6 & 70 \\
	   & M~83	& 04/91    &	  & 1985	& 	& 0.53 	& 400  & 55 & 70$\times$100 & 10 & 45 \\
	   &		& 04/93	   & 	  &  765       	& 	& 0.53  & 300  & & & & \\
	   &		& 12/93	   &	  & 1335	&	& 0.53  & 1000 & & & & \\
$^{12}$CO $J$=4--3   & NGC~6946	& 11/94    & 461  & 8500  	& 11	& 0.51  & 840  & 22 & 30$\times$30  & 6 & 70 \\
	   &		& 07/96    &      & 2900        & 	& 0.53  & 360  & & & & \\
	   & M~83	& 12/93    &      & 4360        &	& 0.51  & 400  & 20 & 30$\times$30  &  6 & 45 \\
\noalign{\smallskip}
\hline
\noalign{\smallskip}
$^{13}$CO $J$=2--1   & NGC~6946    & 02-89 & 220 & 1000 & 21    & 0.63 & 2640  & 2 & & & \\
	   &		& 06-95    &	   	 &  420	&	& 0.69 & 6330  & & & & \\
	   &		& 01-96    &	   	 &  530	&	& 0.69 & 6000 
& & & & \\
	   & M~83	& 02/05-89 &	   	 & 1200	& 21	& 0.63 & 6840  & 3 & & &\\
	   &		& 06-95	   &	   	 &  430	&	& 0.69 & 1200  & & & & \\
$^{13}$CO $J$=3--2   & NGC~6946    & 01-96 & 330 & 2020 & 14    & 0.58 & 6600  & 1 & & & \\
	   & M~83	& 06-00    &	   	 &  644	&	& 0.62 & 2400  & 1 & & & \\
\noalign{\smallskip}
\hline
\noalign{\smallskip}
CI $^{3}$P$_{1}$--$^{3}$P$_{0}$ & NGC~6946 & 11-94 & 492 & 4710 & 10 & 0.43 & 1280 & 17 & 30$\times$24 & 6 & 70 \\
			     &	        & 07-96 &     & 3115 &    & 0.53 &  600 & & & & \\
			     & M~83     & 11-94 &     & 5000 &    & 0.43 &  800 & 14 & 18$\times$36 &  6 & 45 \\
\noalign{\smallskip}
\hline
\end{tabular}
\end{flushleft}
\end{table*}

M~83 (NGC~5236) is likewise a large Sc galaxy. It is part of the
Centaurus A group dominated by the giant elliptical NGC 5128 (the radio
source Cen A) and containing the peculiar galaxies NGC~4945 and NGC~5253 
among others. All main group members have disturbed morphologies 
suggesting recent interactions or mergers. The group contains a large number
of dwarf galaxies (Banks et al. 1999). For M~83, we adopt the group 
distance $D$ = 3.5 Mpc (cf. Israel 1998; Banks et al. 1999). 
Presumably because of its southern declination, M~83 has not been studied
nearly as well as NGC~6946 at (sub)millimeter wavelengths. Early, relatively
low-resolution $J$=1--0 CO measurements were obtained by Rickard et al.
(1977), Combes et al. (1978) and Lord et al. (1987). At a higher resolution
of 16$''$, a $J$=1--0 CO map was published by Handa et al. 1990, showing
a compact central concentration superposed on a `ridge' of CO extending
over 2$'$ in a 45$^\circ$ counterclockwise position angle. Measurements of
the $J$=2--1 and $J$=3--2 transitions of $\co$ and $\13co$ at 22$''$ 
resolution were analyzed by Wall et al. (1993), whereas Petitpas $\&$ 
Wilson (1998) reported on $J$=3--2 and $J$=4--3 CO and 492 GHz
CI maps at similar resolutions. High-resolution aperture synthesis maps
have been published for M~83 in $J$1--0 CO both at the center (Handa et 
al. 1994) and at spiral arm disk positions (Kenney $\&$ Lord 1991; Lord 
$\&$ Kenney 1991; Rand, Lord $\&$ Higdon 1999) as well as in HCN (Helfer 
$\&$ Blitz 1997; Paglione, Jackson $\&$ Ishizuki 1997) -- the center
maps showing a compact, slightly extended source.

\section{Observations}

All observations described in this paper were carried out with the 15m 
James Clerk Maxwell Telescope (JCMT) on Mauna Kea (Hawaii) \footnote{The 
James Clerk Maxwell Telescope is operated on a joint basis between the 
United Kingdom Particle Physics and Astrophysics Council (PPARC), the 
Netherlands Organisation for Scientific Research (NWO) and the National 
Research Council of Canada (NRC).}. Details are given in Table 2.
Up to 1993, we used a 2048 channel AOS backend covering a band of 500 
MHz ($650\kms$ at 230 GHz). After that year, the DAS digital autocorrelator 
system was used in bands of 500 and 750 MHz. Integration times given in 
Table 2 are typical values used in mapping; central positions were usually
observed more than once and thus generally have significantly longer
integration times. Values listed are on+off. When sufficient free baseline was
available, we subtracted second order baselines from the profiles.
In all other cases, linear baseline corrections were applied. All spectra 
were scaled to a main-beam brightness temperature, $T_{\rm mb}$ = 
$T_{\rm A}^{*}$/$\eta _{\rm mb}$; relevant values for $\eta _{\rm mb}$ 
are given in Table 2. Spectra of the central positions in both galaxies are 
shown in Fig. 1 and summarized in Table 3.  In Table 2, we have also listed 
the parameters describing the various maps obtained. All maps are close to 
fully sampled with the exception of the $J$=3--2 CO map of NGC~6946 where 
we sampled the outer parts every other grid point only. In all maps except 
the $J$=2--1 CO map of NGC~6946, the mapping grid was rotated by the angle 
given in Table 2 so that the Y axis coincided with the galaxy major axis. 
The velocity-integrated maps shown in Figs. 2 and 3 have been rotated back, so 
that north is (again) at top and the coordinates are
right ascension and declination. As a consequence of the interpolation 
involved in the rotation, the
maps are shown at a resolution degraded by 5--10$\%$. For NGC~6946, the 
map grid origin is identical to the optical centre listed in Table 1. 
The radio centre occurs in the maps at offsets $\Delta \alpha$, $\Delta 
\delta$ = +2$''$, -1$''$; this is to all practical purposes within the 
pointing error. For M~83, the grid origin is at 13$^{h}$34$^{m}$11.3$^{s}$,
-29$^{\circ}$36$'$39$''$, roughly halfway between the 
optical and radio centres, which occur in the maps at $\Delta \alpha$, 
$\Delta \delta$ = +4$''$, -3$''$ and -3$''$, +4$''$ respectively.

\begin{figure*}
\unitlength1cm
\begin{minipage}[b]{5.75cm}
\resizebox{5.95cm}{!}{\includegraphics*{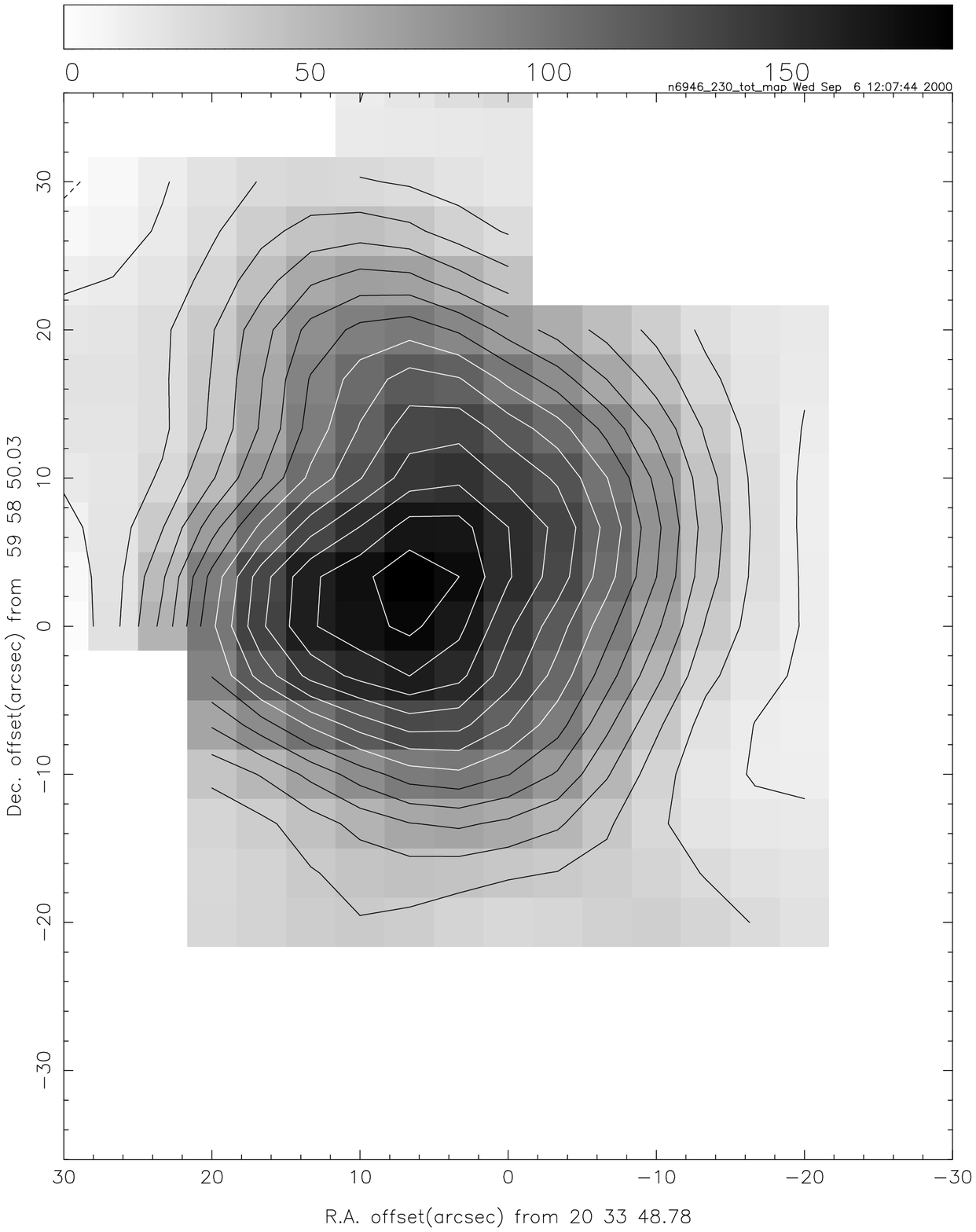}}
\end{minipage}
\hfill
\begin{minipage}[t]{5.75cm}
\resizebox{5.95cm}{!}{\includegraphics*{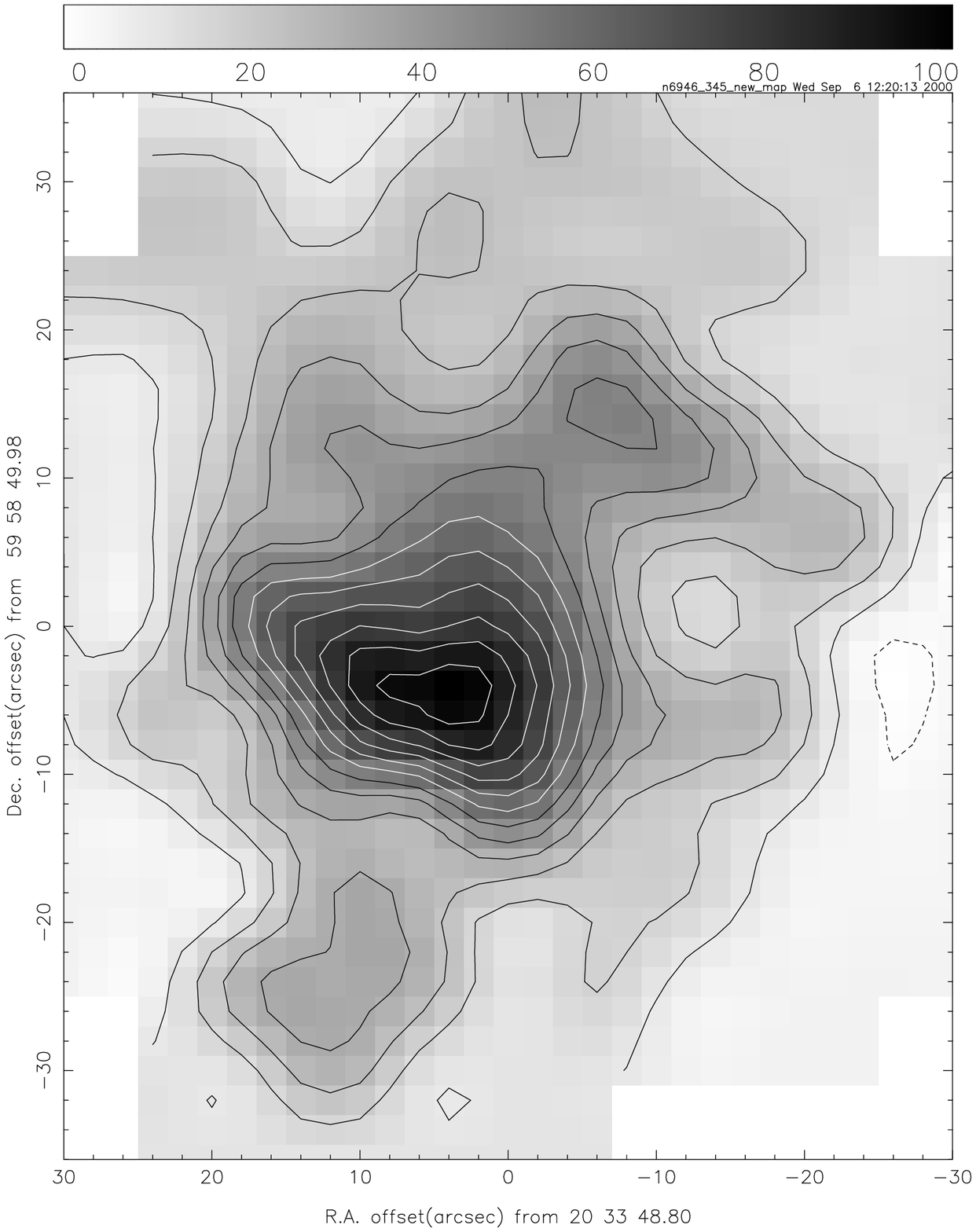}}
\end{minipage}
\hfill
\begin{minipage}[b]{5.75cm}
\resizebox{5.95cm}{!}{\includegraphics*{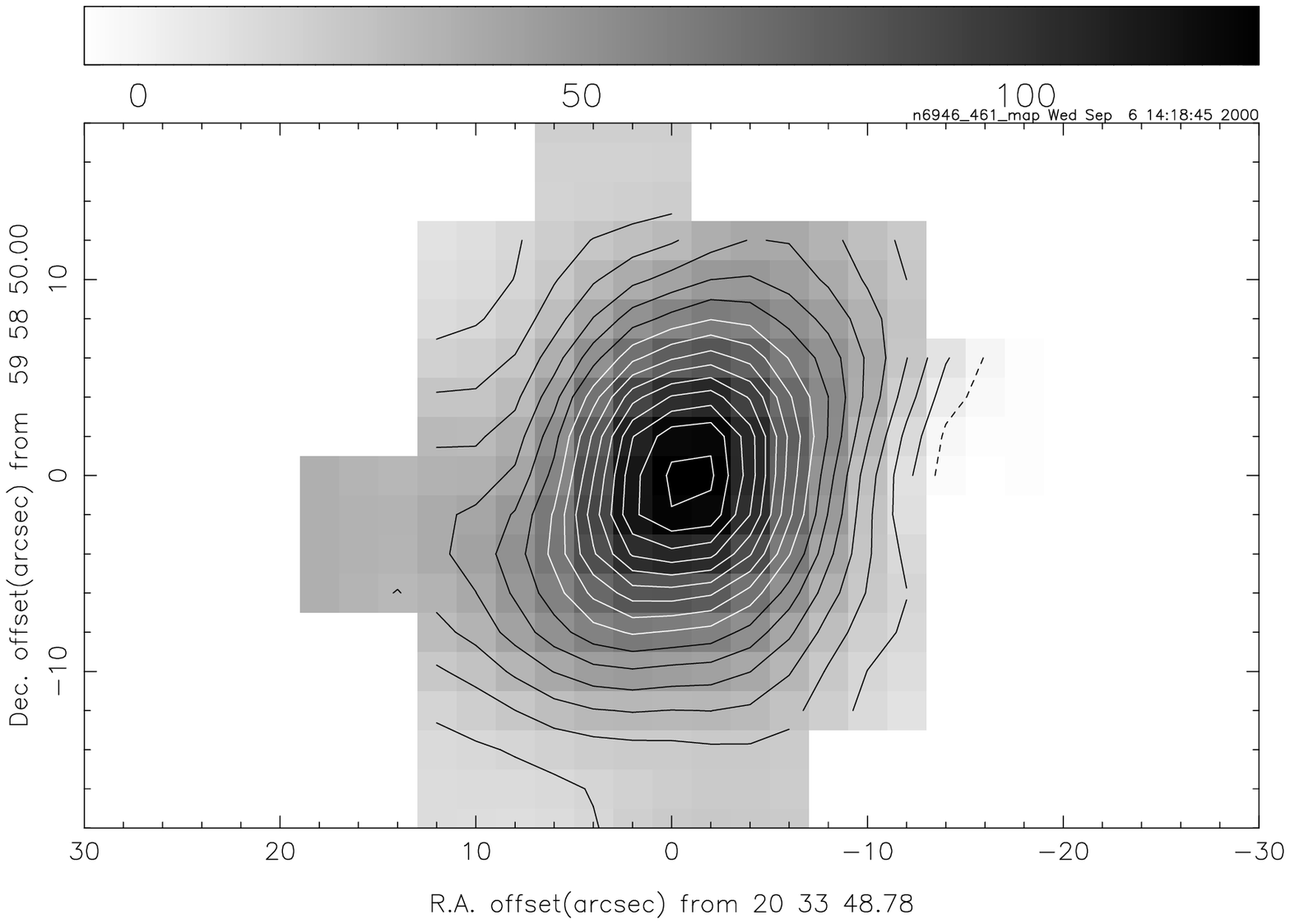}}
\resizebox{5.95cm}{!}{\includegraphics*{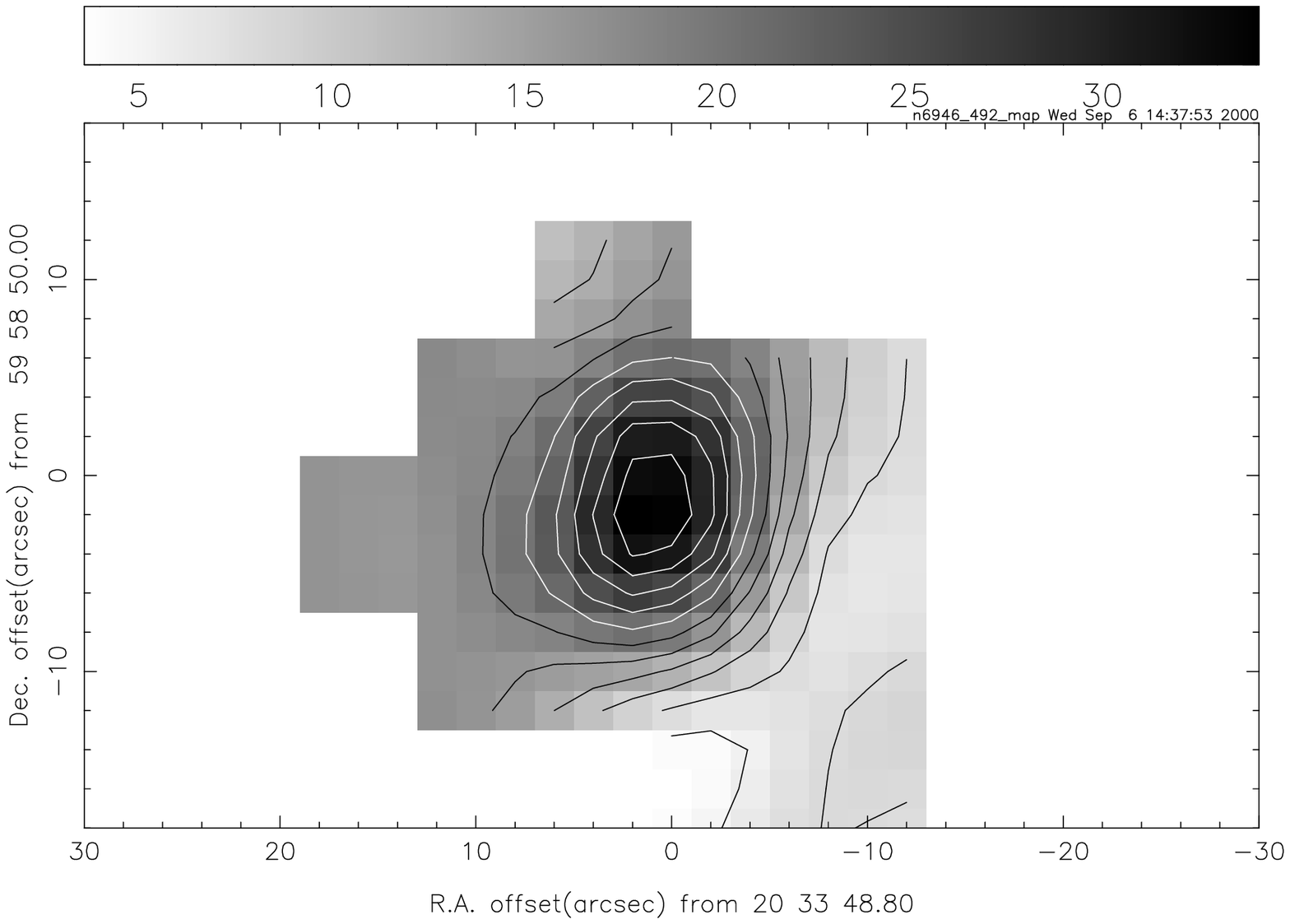}}
\end{minipage}
\caption[]
{Contour maps of emission from NGC~6946 integrated over the velocity
range $V_{LSR}$ = -100 to 220 $\kms$. North is at top. Left to right: 
CO $J$=2--1, CO $J$=3--2, CO $J$=4--3 (top) and [CI] (bottom). Contour 
values are linear in $\int T_{\rm mb} dV$. Contour steps are 20 $\kkms$ 
(2--1), 15 $\kkms$ (3--2 and 4--3) and 6 $\kkms$ (CI) and start at step 1. }
\end{figure*}

\begin{figure*}
\unitlength1cm
\begin{minipage}[b]{5.75cm}
\resizebox{5.95cm}{!}{\includegraphics*{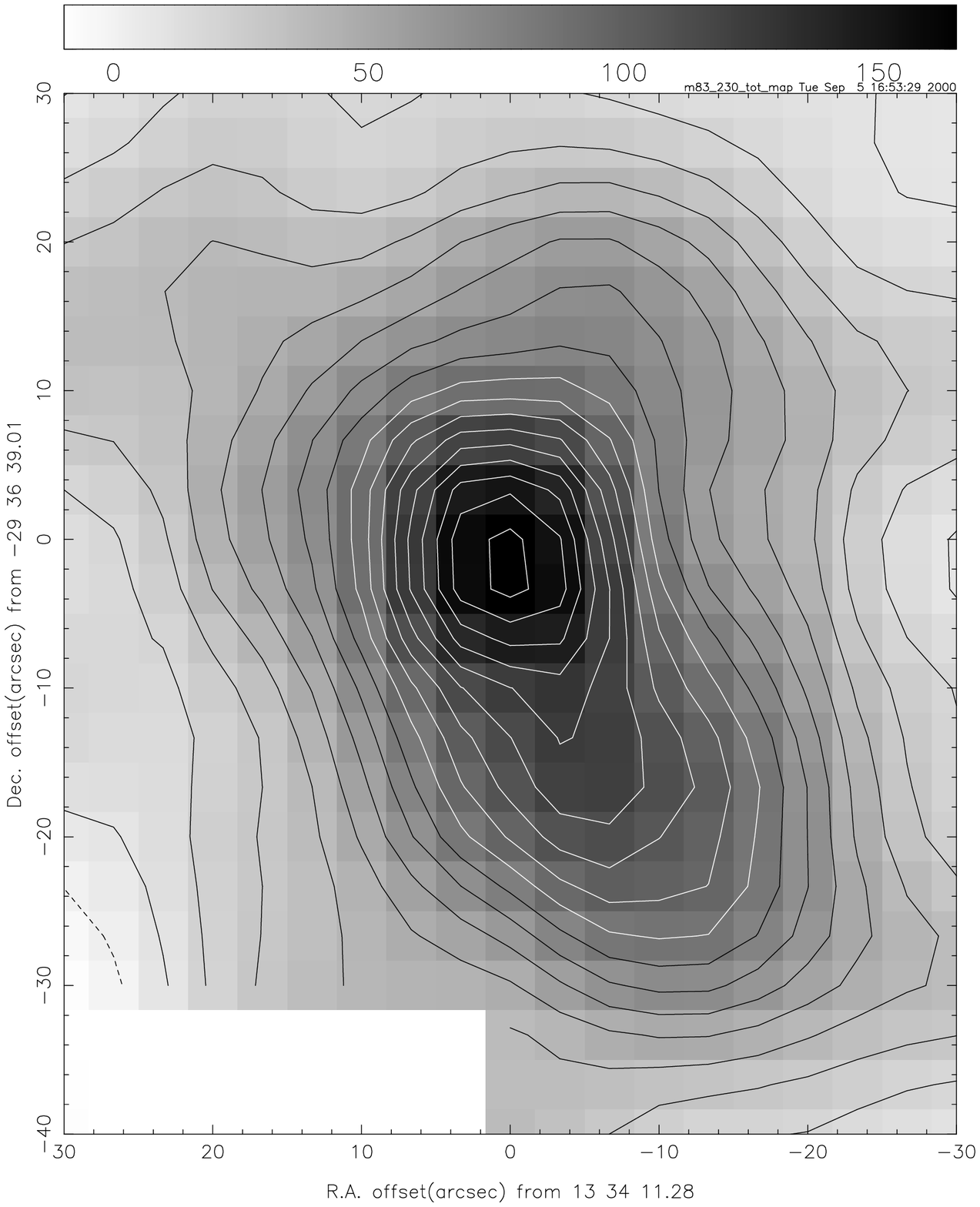}}
\end{minipage}
\hfill
\begin{minipage}[t]{5.75cm}
\resizebox{5.95cm}{!}{\includegraphics*{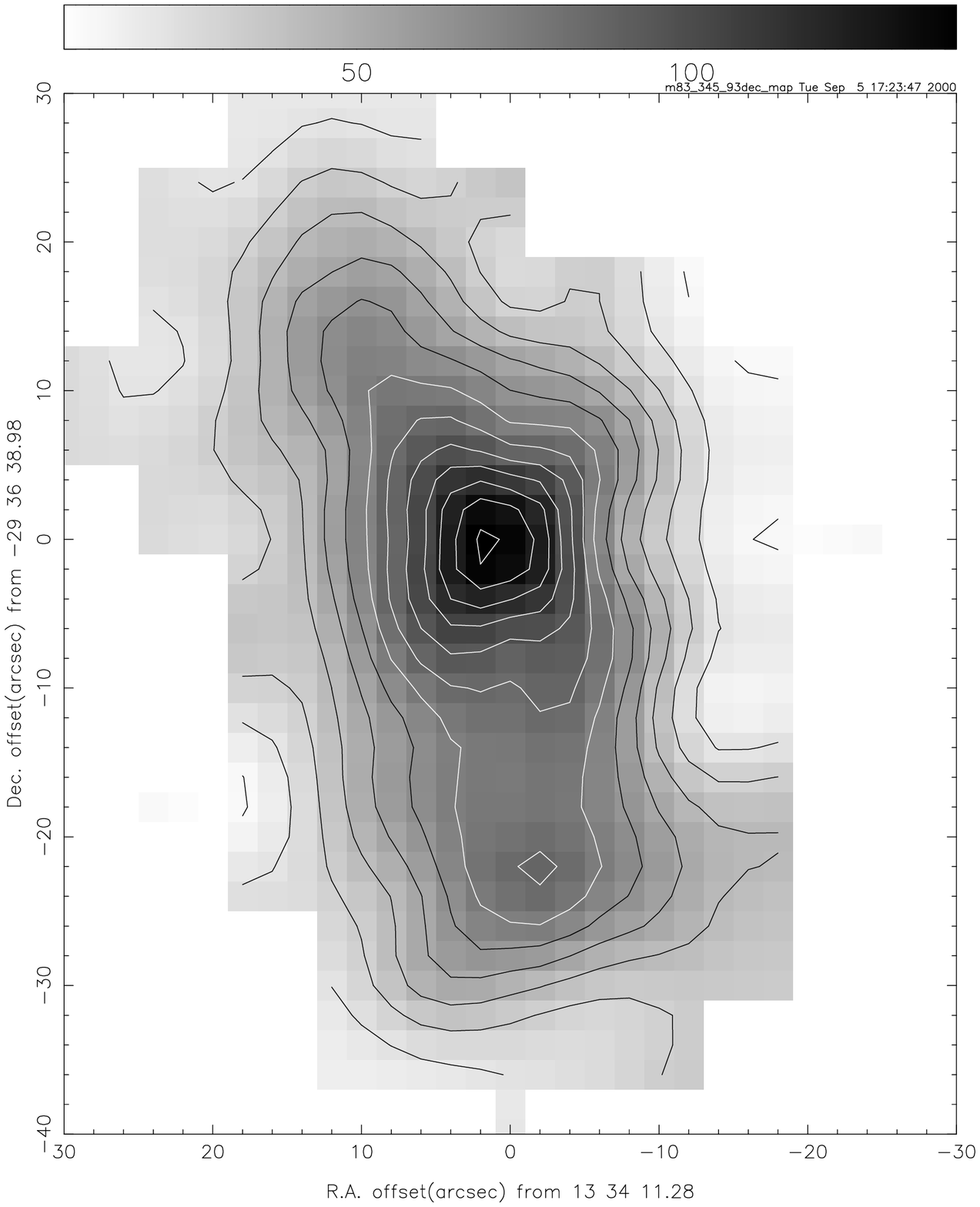}}
\end{minipage}
\hfill
\begin{minipage}[b]{5.75cm}
\resizebox{5.95cm}{!}{\includegraphics*{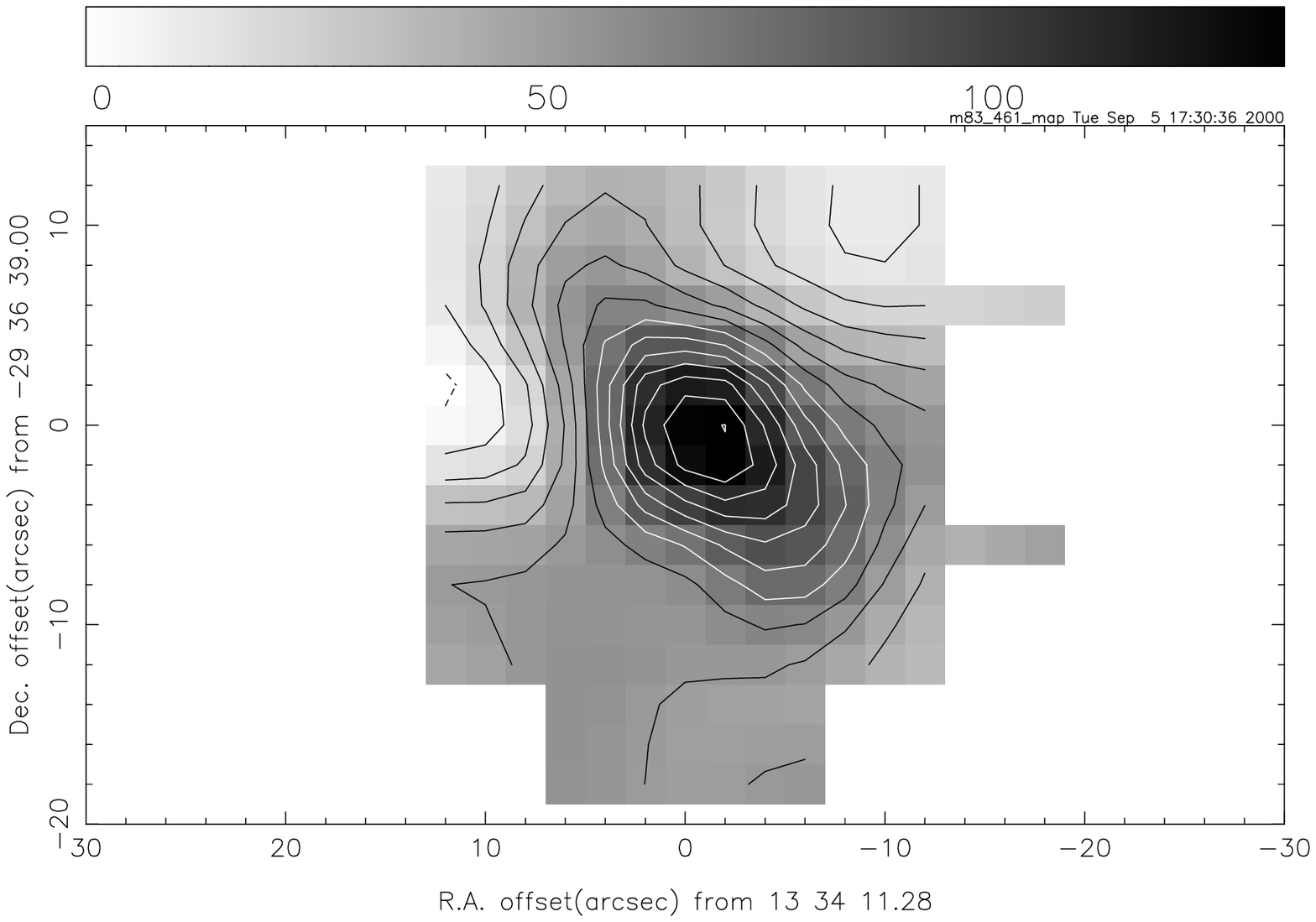}}
\resizebox{5.95cm}{!}{\includegraphics*{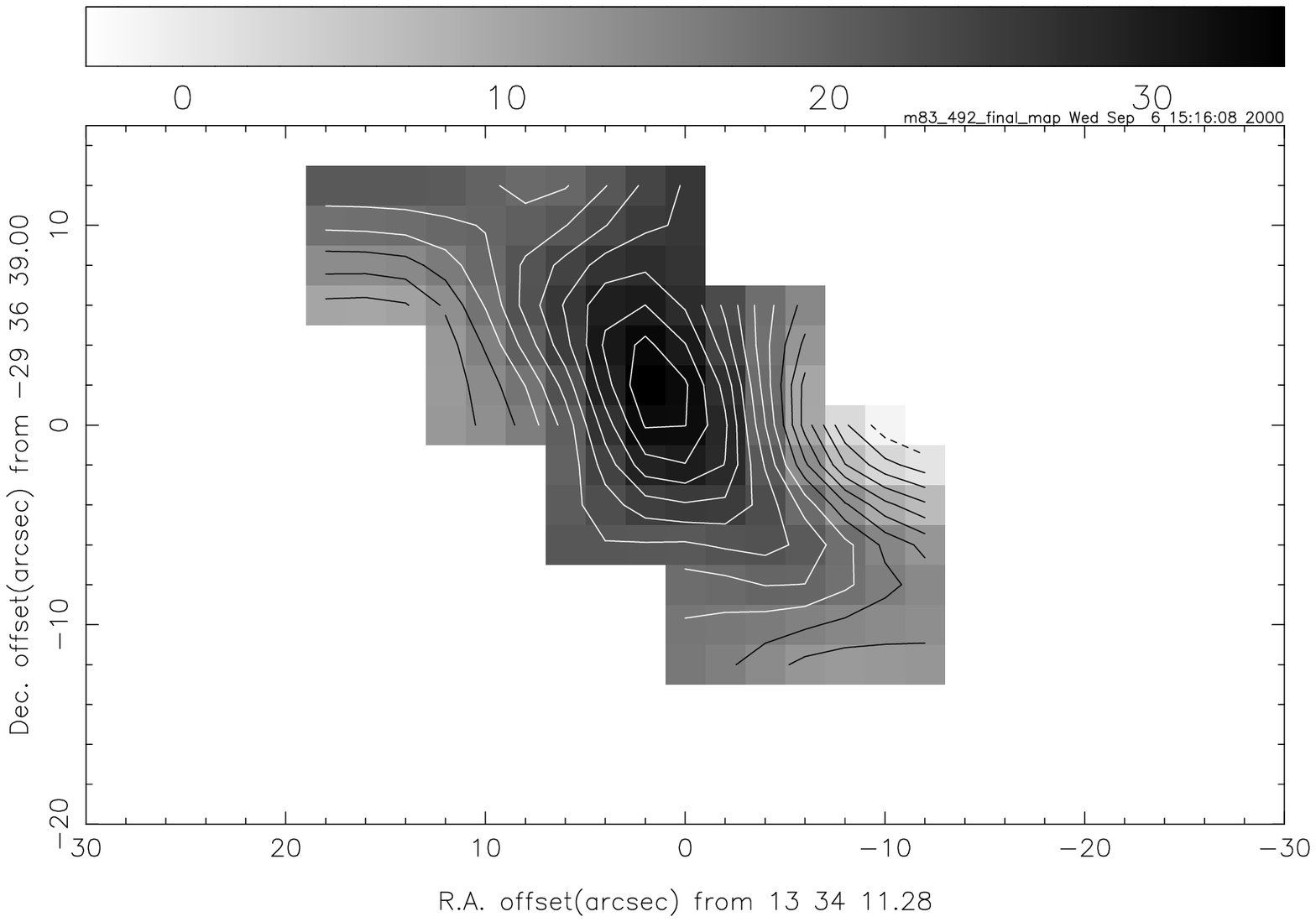}}
\end{minipage}
\caption[]
{Contour maps of emission from M~83 integrated over the velocity range 
$V_{LSR}$ = 400 to 620 $\kms$. North is at top. Left to right: CO $J$=2--1, 
CO $J$=3--2, CO $J$=4--3 (top) and [CI] (bottom). Contour values are 
linear in $\int T_{\rm mb} dV$. Contour steps are 15 $\kkms$ (2--1), 20 
$\kkms$ (3--2 and 4--3) and 5 $\kkms$ (CI) and start at step 1. }
\end{figure*}

\begin{table*}[t]
\caption[]{Central CO and CI line intensities in NGC~6946 and M~83}
\begin{flushleft}
\begin{tabular}{llrrccrcc}
\hline
\noalign{\smallskip}
& & & \multicolumn{3}{l}{NGC~6946} & \multicolumn{3}{l}{M~83} \\
\multicolumn{2}{l}{Transition} & Resolution  
& $T_{\rm mb}$$^a$ & $\int T_{\rm mb}$d$V^a$ & $L_{tot}$$^b$ 
& $T_{\rm mb}$$^a$ & $\int T_{\rm mb}$d$V^a$ & $L_{tot}$$^b$ \\
&       & ($\arcsec$)       & (mK) & ($\kkms$)  & $\kkms$ kpc$^{2}$
			    & (mK) & ($\kkms$)  & $\kkms$ kpc$^{2}$ \\
\noalign{\smallskip}
\hline
\noalign{\smallskip}
$J$=2--1 & $^{12}$CO & 21 & 1384 & 222$\pm$20 & 124 & 2540 & 261$\pm$15 & 55 \\
$J$=3--2 & $^{12}$CO & 14 & 1428 & 209$\pm$25 &  56 & 2176 & 262$\pm$15 & 38 \\
	 &	     & 21 &  	 & 145$\pm$15 &	    &	   & 167$\pm$15 &    \\
$J$=4--3 & $^{12}$CO & 11 & 1798 & 216$\pm$20 &  33 & 2669 & 270$\pm$20 & 21 \\
	 &	     & 14 & 	 & 170$\pm$17 &	    &	   & 189$\pm$20 &    \\
	 &	     & 21 &  	 & 112$\pm$11 &	    &	   & 122$\pm$15 &    \\
\noalign{\smallskip}
$J$=2--1 & $^{13}$CO & 21 &  141 & 22.2$\pm$3 &     &  247 & 28.5$\pm$3 &    \\
$J$=3--2 & $^{13}$CO & 14 &  105 & 11.4$\pm$2 &	    &  194 & 22.3$\pm$1	&    \\
\noalign{\smallskip}
$^{3}$P$_{1}$--$^{3}$P$_{0}$ & CI& 10 & 465 & 85$\pm$9 & 7.2 & 685 & 83$\pm$14 & 4.4 \\
	 &	                  & 14 &     & 60$\pm$10 &   &     & 75$\pm$10  & \\
	 &	                  & 21 &     & 44$\pm$8 &    &     & 55$\pm$8  & \\
\noalign{\smallskip}
\hline
\end{tabular}
\end{flushleft}
Notes to Table 3: a. Beam centered on nucleus; b. Total central concentration
\end{table*}

\begin{table*}
\caption[]{Integrated line ratios in the centres of NGC~6946 and M~83}
\begin{flushleft}
\begin{tabular}{lccccccc}
\hline
\noalign{\smallskip}
Transitions & \multicolumn{3}{l}{NGC~6946} 	 & \multicolumn{4}{l}{M~83} \\
	    & Nucleus & Total Center & +10$''$, +10$''$  & Nucleus & Total Center & +7$''$, -7$''$ & -14$''$, -14$''$ \\ 
\noalign{\smallskip}
\hline
\noalign{\smallskip}
$\co$ (1--0)/(2--1)$^{a}$    & 1.1$\pm$0.2   & 0.95 & 1.0  & 0.9$\pm$0.2    & 1.1  & ---  & ---  \\
$\co$ (3--2)/(2--1)$^{b}$    & 0.65$\pm$0.10 & 0.5  & 0.5  & 0.65$\pm$0.13  & 0.7  & 0.8  & 0.5: \\
$\co$ (4--3)/(2--1)$^{b}$    & 0.45$\pm$0.15 & 0.3  & 0.4: & 0.48$\pm$0.10  & 0.4  & 0.6  & ---  \\
\noalign{\smallskip}
$\co$/$\13co$ (1--0)$^{c}$ & 11.8$\pm$1.3  & ---  & ---  & 10.4$\pm$1.6   & ---  & ---  & ---  \\
$\co$/$\13co$ (2--1)$^{b}$ &  9.8$\pm$1.5  & ---  & 15   &  9.2$\pm$1.0   & ---  & 11   & 9    \\
$\co$/$\13co$ (3--2)$^{d}$ & 13.0$\pm$1.4  & ---  & ---  & 11.8$\pm$1.1   & ---  & ---  & ---  \\
\noalign{\smallskip}
CI/CO(2--1)$^{b}$   		 & 0.20$\pm$0.04 & 0.06 & ---  & 0.18$\pm$0.04  & 0.07 & 0.2: & ---  \\
CII/CO(2--1)$^{e}$  		 & 0.08	         &	&      & 0.55	    	&      &      &      \\
\noalign{\smallskip}
\hline
\end{tabular}
\end{flushleft}
Notes: 
a. From $J$=1--0 data by Weliachew et al. (1988), Sofue et al. 
   (1988), Wild (1990); Handa et al. (1990) and Israel et al. (unpublished);
b. This Paper, JCMT at 21$''$ resolution;
c. Sage $\&$ Isbell (1991); NRAO 12m at 57$''$ resolution; 
   Young $\&$ Sanders (1986); FCRAO at 45$''$ resolution; Israel et al 
   (unpublished); SEST at 43$''$ resolution; Rickard $\&$ Blitz (1985); 
   NRAO at 65$''$ resolution;
d. This Paper; JCMT at 14$''$ resolution.
e. From Crawford et al. (1985) and Stacey et al. (1991), KAO at 55$''$ 
   resolution.
\end{table*}

\begin{figure*}
\unitlength1cm
\begin{minipage}[b]{5.54cm}
\resizebox{5.8cm}{!}{\includegraphics*{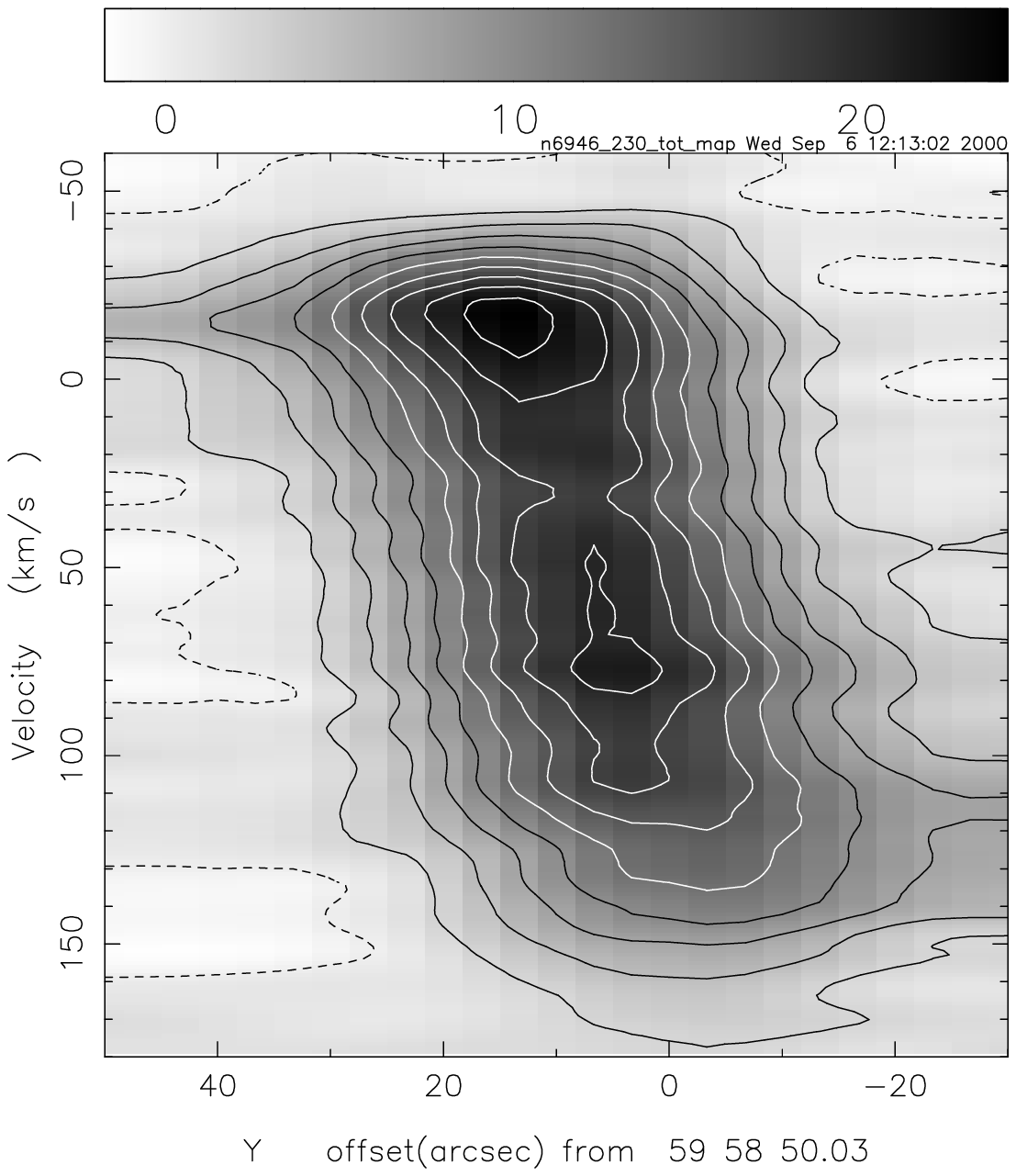}}
\end{minipage}
\hfill
\begin{minipage}[t]{5.54cm}
\resizebox{5.8cm}{!}{\includegraphics*{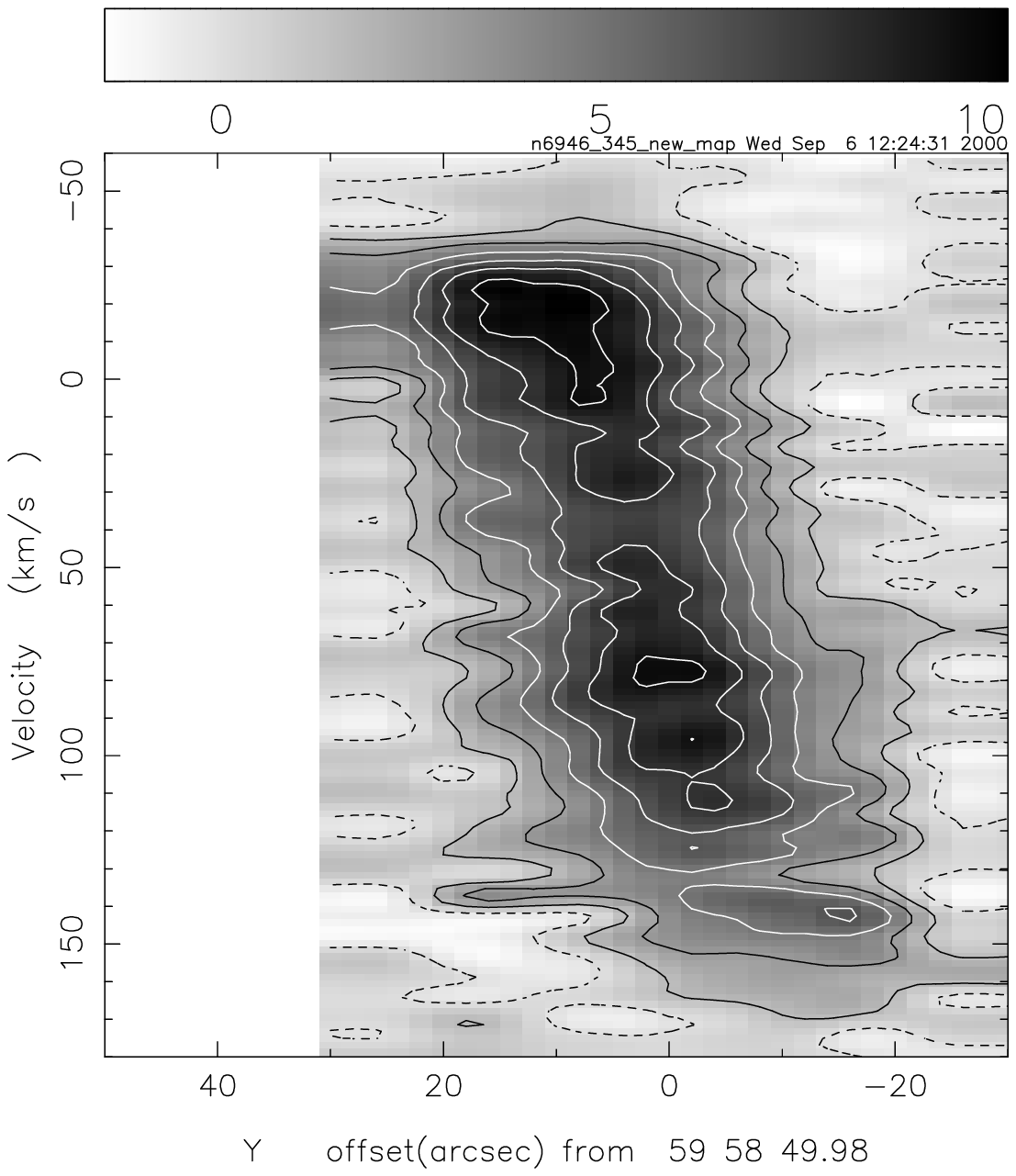}}
\end{minipage}
\hfill
\begin{minipage}[t]{5.54cm}
\resizebox{5.8cm}{!}{\includegraphics*{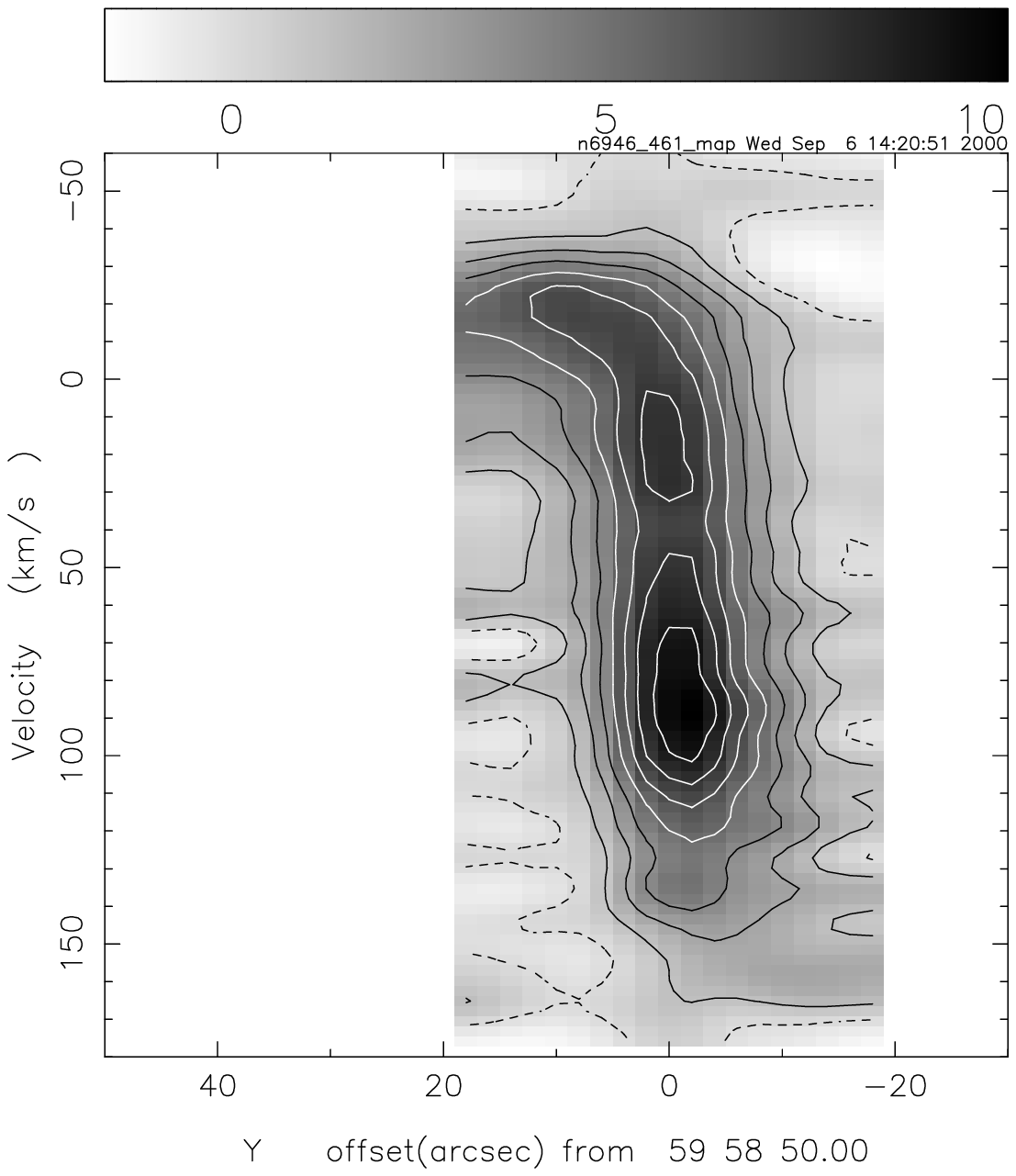}}
\end{minipage}
\caption[]
{Position-velocity maps of CO emission from NGC~6946 in position angle
55$^{\circ}$. Left to right: CO $J$=2--1, CO $J$=3--2 and CO $J$=4--3. 
Contour values are linear in $T_{\rm mb}$. Contour steps 4 K (2--1), 
3 K (3--2) and 2.5 K (4--3) and start at step 1.
}
\end{figure*}

\begin{figure*}
\unitlength1cm
\begin{minipage}[b]{5.54cm}
\resizebox{6cm}{!}{\includegraphics*{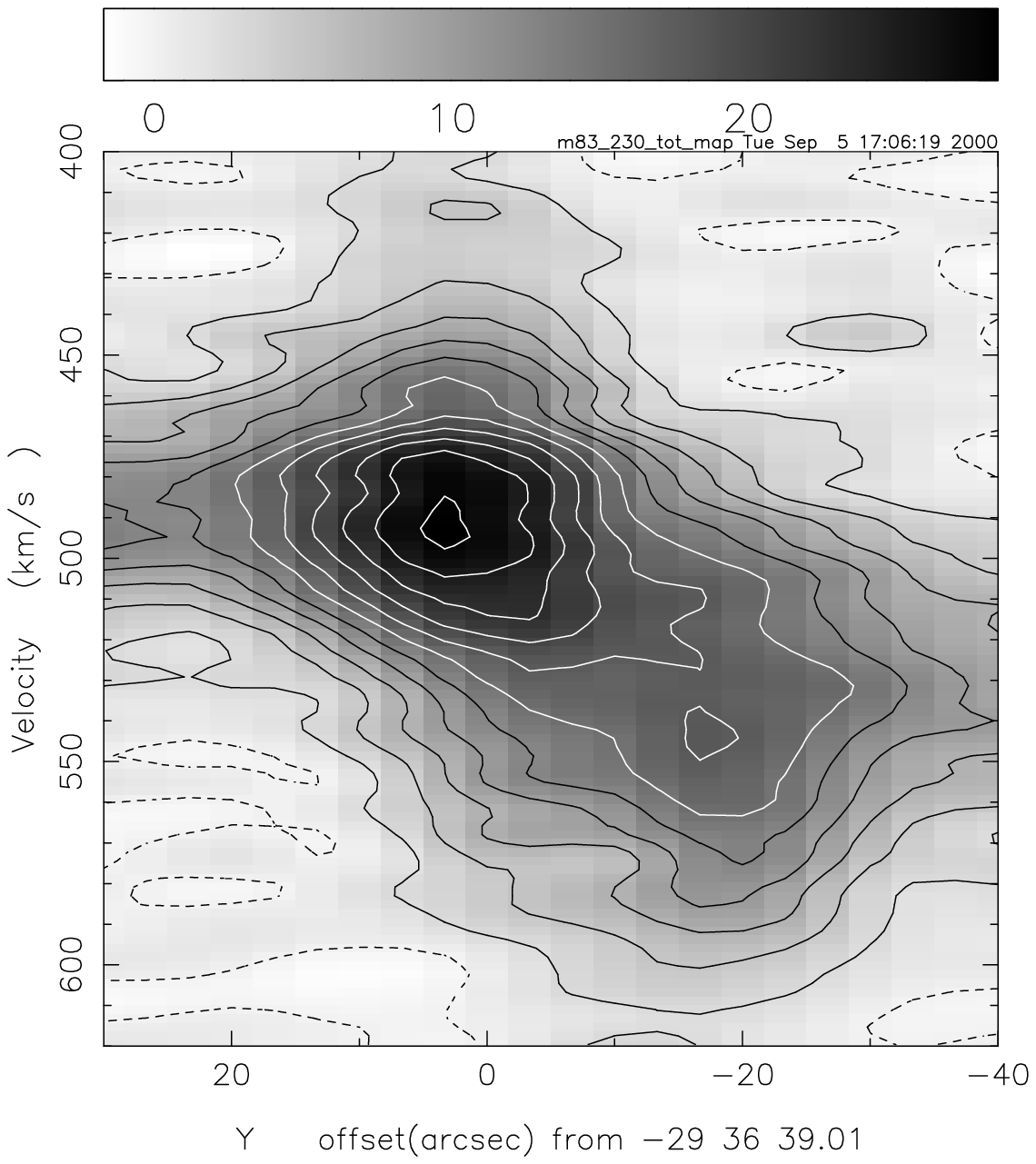}}
\end{minipage}
\hfill
\begin{minipage}[t]{5.54cm}
\resizebox{6cm}{!}{\includegraphics*{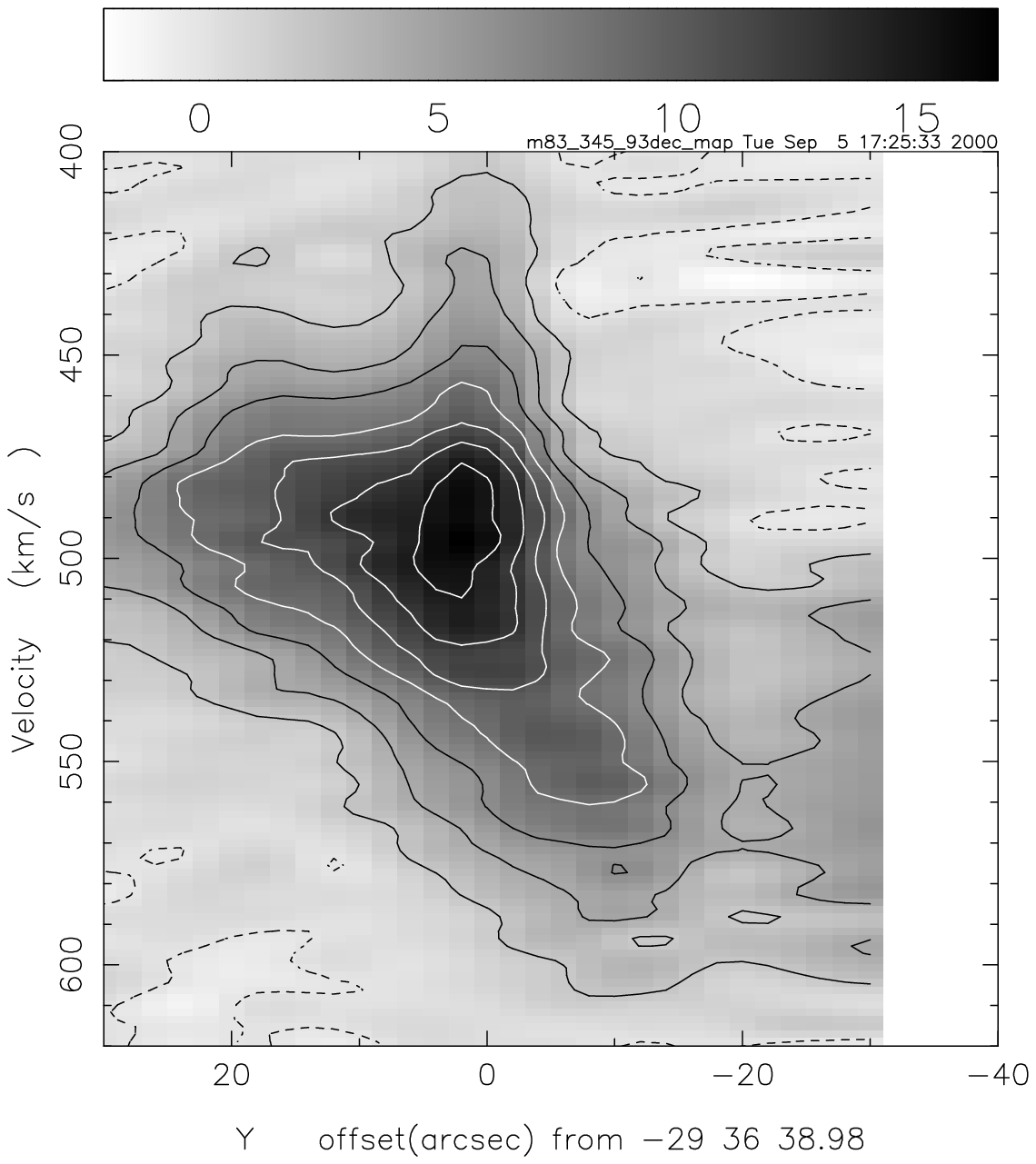}}
\end{minipage}
\hfill
\begin{minipage}[t]{5.54cm}
\resizebox{6cm}{!}{\includegraphics*{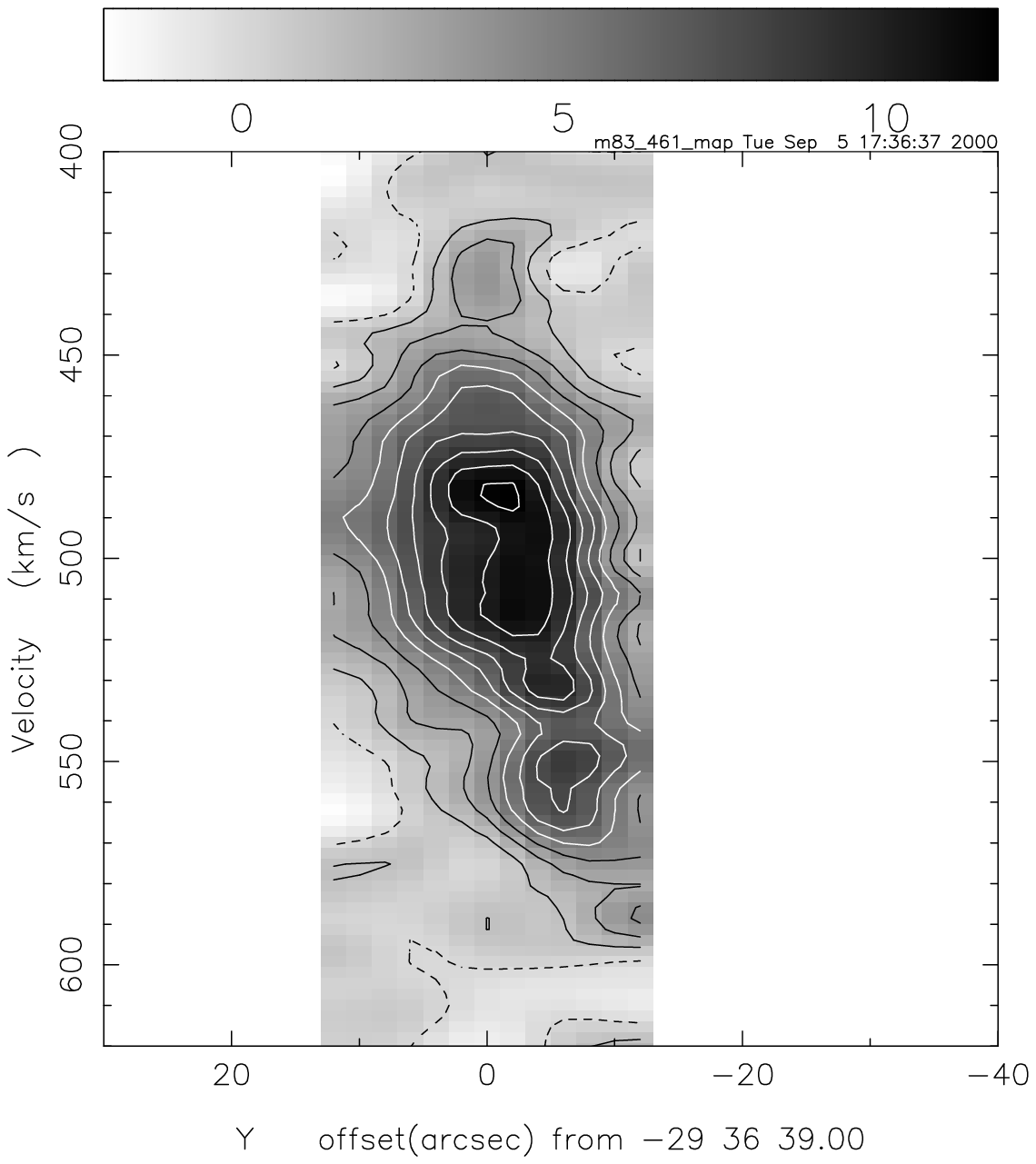}}
\end{minipage}
\caption[]
{Position-velocity maps of CO emission from M~83 in position angle 
45$^{\circ}$). Left to right: CO $J$=2--1, CO $J$=3--2 and CO $J$=4--3. 
Contour values are linear in $T_{\rm mb}$. Contour steps are 4 K (2--1 and 
3--2) and 2.5 K (4--3) and start at step 1.
}
\end{figure*}

\section{Results}

\subsection{CO distribution}

In both galaxies, there is a strong concentration of molecular material
in the central region. The central source, although not dominating
the {\it total} CO emission from the galaxy, is nevertheless a major
feature compared with the minor peaks occurring in the disk (see
the NGC~6946 CO maps by Casoli et al. 1990 and Sauty et al. 1998, as
well as the M~83 interferometer map by Rand et al. 1999). A similar 
impression is provided by the SCUBA 850$\mu$m continuum map of NGC~6946 
(Bianchi et al. 2000), although the continuum image of the central 
source in particular is seriously contaminated by $J$=3--2 CO line 
emission. 

In NGC~6946, the strong central CO emission is highly structured,
as revealed in the $J$=1--0 interferometer map by Regan $\&$ Vogel 
(1995), and also partly in our lower-resolution maps in Fig. 2, 
in particular in the $J$=3--2 map, which shows close resemblance
to their $J$=1--0 map. The central region of NGC~6946 has very similar
CO and optical morphologies (Regan $\&$ Vogel 1995, see also 
Ables 1971). The maps show strong centralized emission 
superposed on more extended emission of lower surface brightness. 
The overall extent of the central CO source in NGC~6946 is about 
$50''\times25''$. Most of the extended emission occurs roughly along 
the minor axis of the galaxy and appears to be due to enhanced CO emission 
from spiral arm segments (cf. Regan $\&$ Vogel 1995) out to about 
$R$ = 1 kpc in the plane of the galaxy. In addition to these minor axis
extensions, there are also extensions along the major axis, particularly 
in the ENE direction. The bright central peak is especially prominent 
in the $J$=4--3 CO and [CI] maps, its increased contrast in these maps
being caused mainly by higher resolution and higher excitation (see 
below). The source extent of about 10$''$ in these maps is 
consistent with the $J$=1--0 CO scale length $r_{e}$ = 160 pc derived
by Sakamoto et al. (1999). This compact source has also been detected
and mapped interferometrically in $J$=1--0 HCN (Helfer $\&$ Blitz 1997). 
Evidence for further, unresolved structure is provided by the central 
emission profiles in Fig. 1 and the major axis position-velocity
maps in Fig. 4. They show a clear double-peaked structure in all 
transitions with a minimum at about $V_{LSR}$ = +45 $\kms$ suggesting 
a deficit of material (a `hole') at the very center of NGC~6946. 
The position-velocity maps show a steep central velocity gradient,
undiscernible from rapid solid-body rotation, with steepness 
apparently increasing with increasing $J$ number. The significantly 
lesser steepness in e.g. the $J$=2--1 CO map is caused by beamsmearing. 
This is readily seen from a comparison of the $J$=4--3 CO (Fig. 4) and 
the high-resolution $J$=1--0 CO (Fig. 4 in Sakamoto et al. 1999) 
velocity gradients which are practically identical with d$V$/d$\theta$ 
= 35 $\kms$/$''$ (in the plane of the galaxy corresponding to d$V$/d$R 
\approx$ 2 $\kms$/pc). From this gradient and the velocity separation 
of the central profile peaks in Fig. 1, we estimate the size of the 
`hole' in the disk to be of the order of 2$''$ ($R$ = 25 pc). The steep 
rotation curve turns over to a much flatter one at a radius of about $R$ 
= 200 pc.

The structure of the central CO source in M~83 is, at least with the
presently available data, much simpler. A central peak resolved only 
at resolutions $\leq$ 15$''$ is superposed on a more extended 
ridge along the major axis seen in the $J$=3--2, $J$=2--1 and $J$=1--0 
CO maps (Fig. 3; see also Handa et al. 1990) with dimensions $55''\times25''$. 
The ridge thus extends outwards to a radius of about $R$ = 1 kpc, so
that the overall sizes of the central CO source in M~83 and NGC~6946 are 
very similar. The ridge shows some structure, perhaps including two 
symmetrical secondary maxima each at about $R$ = 325 pc from the nucleus. 
In the $J$=4--3 CO and [CI] maps (Fig. 3), the central peak is just resolved 
along the major axis, extending to a radius $R$ = 135 pc from the nucleus. 
Along the minor axis, it is unresolved. As is the case with NGC~6946, 
the contrast of the peak with its surroundings is higher than that in
lower $J$ transitions, at least in part because of higher resolution.
Compact, barely resolved emission from the peak is also seen in the 
$J$=1--0 HCN transition interferometrically mapped by Helfer $\&$ Blitz 
(1997). The central emission profiles (Fig. 1 bottom) of M~83 do not resemble
those of NGC~6946. They are clearly non-gaussian, but instead of a 
double-peaked shape, they are perhaps best described as a slightly
asymmetric blend of a broad and a narrow component.

The position-velocity maps in Fig. 5 are quite interesting. A compact 
component in very rapid solid-body rotation is shown superposed on 
more extended emission in much more sedate rotation. The effect of
beamsmearing is particularly noticeable in the apparently much greater
extent of the rapidly rotating component in the lower $J$ maps (for
$J$=1--0 CO, see Handa et al. 1990). From the $J$=4--3 CO map in Fig. 5,
we find that the rapidly rotating disk is contained with $R$ = 95 pc from
the nucleus, and that it has a velocity gradient  d$V$/d$\theta$ = 18 
$\kms$/$''$, corresponding to d$V$/d$R$ = 2.7 $\kms$/pc in the plane of
the galaxy. The more extended material has a velocity gradient d$V$/d$\theta$ 
= 0.6 $\kms$/$''$, corresponding to only d$V$/d$R$ = 85 $\kms$/kpc in the 
plane of the galaxy. The $J$=4--3 CO and [CI] maps suggest that the rapidly
rotating material is a relatively thin disk. Our results do not
provide any evidence for the presence of the small central hole that may be
surmised from the aperture synthesis observations by Handa et al. (1994).
The JCMT $J$=3--2, $J$=4--3 and [CI] maps published by Petitpas $\&$
Wilson (1997) show a clear double-peaked structure, the peaks being 
separated by some 16$''$. Our $J$=3--2 and $J$=4--3 CO maps do not
reproduce the structure seen by Petitpas $\&$ Wilson (1997). In 
particular they do not show the secondary peak which should occur at 
$\Delta \alpha, \Delta \delta$ = +4$''$, -9$''$ in our maps; that position 
is not fully covered by either [CI] map. Signal-to-noise ratio considerations,
the poor baselines encountered by Petitpas $\&$ Wilson (1997) in the their
$J$=3--2 observations and the limited extent (about one beamwidth) of the 
secondary peak lead us to question its prominence and perhaps even its
existence.  

\subsection{Line ratios}

From our observations, we have determined the intensity ratio of the 
observed transitions at various positions in both galaxies. For 
convenience, we have normalized all intensities to that of the $J$=2--1 
$\co$ line. All $^{12}$CO ratios given for individual positions refer to a 
beam of 21$''$; where necessary we convolved higher-resolution observations to 
this beamsize to obtain an accurate ratio not susceptible to varying 
degrees of beam dilution. Isotopic $^{12}$CO/$^{13}$CO ratios are given for 
the resolutions listed in the Table. The individual positions include, in 
addition to the nuclear positions of both galaxies, an offset position in 
NGC~6946 representing the off-nucleus CO cloud complex discussed in sect. 
4.3, and two offset positions in M~83 on the major and minor axis respectively.
The $J$=1--0/$J$=2--1 ratios have relatively large uncertainties,
because we have used $J$=1--0 intensities estimated for a 21$''$ beam 
from the references given in the table. These ratios are, in any case, 
close to unity. 

In contrast, the columns in Table 4 marked Total Center refer to the 
intensities integrated over total source extent as shown in the
maps. At the lower frequencies, source extents are larger than at the
higher frequencies. This is mostly caused by limited and frequency-dependent
resolution. When corrected for finite beamwidth, source dimensions at e.g. 
the $J$=2--1 and $J$=4--3 transitions are very similar for NGC 6946 and the 
bright peak of M~83. Nevertheless, the smaller area coverage at the
higher frequencies may lead to an underestimate of the intensities of 
the emission at these frequencies and consequently the corresponding 
line ratios especially if extended emission of relatively low surface 
brightness is present. The entries in Table 4 suggest that this may 
indeed be the case for $J$=4--3 CO and [CI].

We have converted the [CII] intensities measured by Crawford et al. (1985) 
and Stacey et al. (1991) to velocity-integrated temperatures. The line 
ratios given in Table 4 were obtained after convolving our $J$=2--1 CO 
results to the same beam solid angle of 8.6 $\times$ 10$^{-8}$ sr 
(HPBW 55$''$) that was used to measure the [CII]. 

It is quite remarkable that NGC~6946 and M~83 are extremely similar in
all ratios (and indeed intensities), except for the [CII] intensity. 
From the observed CO transitions only it is easily but mistakenly 
concluded that the two galaxies have identical ISM properties in their 
center. As it is, the intensity of the [CII] line suggests a much stronger
PDR effect in M~83 than in NGC~6946, implying the presence of both high 
gas temperatures and densities in the medium as the critical values for 
this transition are $T_{\rm kin} \geq$ 91 K and $n \geq 3500 \cc$. At the 
same time, such values must be reconciled with the much lower temperatures 
and (column) densities implied by the modest CO isotopic ratios.

\section{Analysis}

\subsection{Modelling}

\begin{table*}
\caption[]{Model parameters}
\begin{flushleft}
\begin{tabular}{lccccccccc}
\hline
\noalign{\smallskip} 
Model & \multicolumn{3}{c}{Component 1}    	 & \multicolumn{3}{c}{Component 2}  	     & Ratio$^{a}$ & \multicolumn{2}{c}{Line Ratios} \\
     & $T_{\rm k}$  & $n(H_{2}$) & $N(CO)$/d$V$	 & $T_{\rm k}$ 	& $n(H_{2}$) & $N(CO)$/d$V$  & Comp. 	   & $\co^{b}$ & $\13co^{c}$ \\
     & (K)     	    & ($\cc$)    & ($\cm2/\kms$) & (K) 		& ($\cc$     & ($\cm2/\kms$) & 1:2   & & \\
\noalign{\smallskip}
\hline
\noalign{\smallskip}
\multicolumn{10}{c}{NGC~6946}\\
\noalign{\smallskip}
\hline
\noalign{\smallskip}
1 &  30 & 1000 & 10$\times10^{17}$ & 150 & 1000   & 0.3$\times10^{17}$ & 1:9 & 1.22 0.69 0.40 & 11  9.9 13 \\
2 &  60 & 1000 &  1$\times10^{17}$ &  30 & 10 000 & 0.6$\times10^{17}$ & 6:4 & 1.12 0.73 0.41 & 11 10.0 13 \\
3 & 150 &  500 &  1$\times10^{17}$ &  30 & 10 000 & 0.6$\times10^{17}$ & 8:2 & 1.18 0.73 0.44 & 11  9.8 13 \\ 
4 & 100 & 1000 &  1$\times10^{17}$ & --- &  ---   & ---                & --- & 1.33 0.73 0.44 & 12  9.3 13 \\
\noalign{\smallskip}
\hline
\noalign{\smallskip}
\multicolumn{10}{c}{M~83} \\
\noalign{\smallskip}
\hline
\noalign{\smallskip}
5 &  30 & 3000 &  1$\times10^{17}$ & 100 & 3000   & 0.06$\times10^{17}$ & 4:6 & 0.93 0.73 0.40 & 11 8.9 12 \\
6 &  60 & 1000 &  1$\times10^{17}$ &  60 & 100000 &    1$\times10^{17}$ & 9:1 & 1.14 0.76 0.49 & 10 9.5 12 \\
7 & 150 &  500 &  3$\times10^{17}$ &  60 & 3000   &  0.6$\times10^{17}$ & 3:7 & 1.13 0.77 0.51 & 10 9.0 12 \\
8 & 100 & 1000 &  1$\times10^{17}$ & --- & ---	  &	---		& --- & 1.33 0.73 0.44 & 12 9.3 13 \\
\noalign{\smallskip}
\hline
\end{tabular}
\end{flushleft}
Notes $^{a}$ Ratio denotes the relative contributions of the two 
components to the observed emission in the $J$=2--1 $\co$ line.\\
$^{b}$ Model-calculated intensities of the
$J$=1--0, $J$=3--2 and $J$=4--3 $\co$ transitions normalized to
the $J$=2--1 $\co$ intensity.\\
$^{c}$ $\co/\13co$ intensity ratios in the $J$=1--0, $J$=2--1 and 
$J$=3--2 transitions. 
\end{table*}

\begin{table*}
\caption[]{Beam-averaged results}
\begin{flushleft}
\begin{tabular}{lccccccc}
\hline
\noalign{\smallskip} 
Model & \multicolumn{3}{c}{Beam-Averaged Column Densities} & \multicolumn{2}{c}{Total Central Mass} & \multicolumn{2}{c}{Face-on Mass Density} \\
   & $N(CO)$ & $N(C)$    & $N({\it \h2})$ & $M(\h2)$  & $M_{\rm gas}$        & $\sigma(\h2)$ & $\sigma_{\rm gas}$ \\
   & \multicolumn{2}{c}{($10^{18} \cm2$)} & ($10^{21} \cm2$) & \multicolumn{2}{c}{($10^{7}$ M$_{\odot}$)} & \multicolumn{2}{c}{(M$_{\odot}$/pc$^{-2}$)} \\
\noalign{\smallskip}
\hline
\noalign{\smallskip}
\multicolumn{8}{c}{NGC~6946; $N_{H}/N_{C}$ = 2500; $N(HI)^{a} = 1.3 \times 10^{21} \cm2$} \\
\noalign{\smallskip}
\hline
\noalign{\smallskip}
1  & 1.5 & 0.9 &  2.4 & 2.1 & 3.6 &  29 &  51 \\
2  & 0.7 & 1.3 &  2.1 & 1.8 & 3.3 &  26 &  45 \\
3  & 0.9 & 1.2 &  2.0 & 1.8 & 3.2 &  25 &  45 \\
4  & 0.8 & 1.1 &  1.7 & 1.5 & 2.8 &  21 &  39 \\
\noalign{\smallskip}
\hline
\noalign{\smallskip}
\multicolumn{8}{c}{M~83; $N_{H}/N_{C}$ = 2500; $N(HI)^{b} = 0.6 \times 10^{21} \cm2$} \\
\noalign{\smallskip}
\hline
\noalign{\smallskip}
5  & 0.8 & 3.9 &  5.5 & 1.8 & 2.6 &  80 & 114 \\
6  & 1.0 & 4.3 &  6.3 & 2.4 & 3.4 &  92 & 130 \\
7  & 1.1 & 3.7 &  5.7 & 2.1 & 3.0 &  83 & 118 \\
8  & 0.9 & 8.2 & 11.0 & 3.7 & 5.1 & 159 & 221 \\
\noalign{\smallskip}
\hline
\end{tabular}
\end{flushleft}
Notes: a. Boulanger $\&$ Viallefond (1992); b. Rogstad et al. (1974); 
Tilanus $\&$ Allen (1993)
\end{table*}

The observed $\co$ and $\13co$ line intensities and ratios can be modelled 
by radiative transfer models such as described by Jansen (1995) and
Jansen et al. (1994). The models provide line intensities as a function
of three input parameters: gas kinetic temperature $T_{\rm k}$, molecular 
hydrogen density $n(H_{2})$ and CO column density per unit velocity 
($N(CO)$/d$V$). By comparing model line {\it ratios} to the observed ratios we 
may identify the physical parameters best describing the actual conditions in 
the observed source. The additional filling factor is found by comparing model 
{\it intensities} to those observed. If $\co$ and $\13co$ have the same beam
filling factor, a single component fit requires determination of five
independent observables. As we have measured seven line intensities, such 
a fit is, in principle, overdetermined. In practice, this is not quite the 
case because of significant finite errors in observed intensities, and 
because of various degrees of degeneracy in the model line ratios. We found 
that a single-component fit could be made to the data of the two galaxies 
only if we allow CO $J$=1--0 intensities to be rather higher than observed. 
As we also consider a single temperature, single density gas to be 
physically implausible for the large volumes sampled, we reject such a fit.

It is much more probable that the large linear beams used sample molecular
gas with a range of temperatures and densities. We approximate such
a situation by assuming the presence of two independent components.
As this already doubles the number of parameters to be determined to ten, 
a physically realistic more complex analysis is not possible. In our 
analysis, we assume identical 
CO isotopical abundances for both components, and by assuming a specific 
value (i.e. [$^{12}$CO]/[$^{13}$CO] = 40, cf. Mauersberger $\&$ Henkel 
1993) reduce the number of parameters to eight. This leaves a single free 
parameter, for which we take the relative contribution (filling factor) of 
the two components in the emission from the $J$=2--1 $\co$ line.  
Acceptable fits are then identified by searching a grid of model
parameter combinations (10 K $\leq T_{\rm k} \leq $ 250 K, $10^{2} \cc \leq
n(\h2) \leq 10^{5} \cc$, $6 \times 10^{15} \cm2 \leq N(CO)/dV \leq 
3 \times 10^{18} \cm2$; relative emission contributions 0.1 to 0.9) for sets 
of line ratios matching the observed set.

The CO line ratios observed for NGC~6946 and M~83 can be fit by various 
combinations of gas parameters or rather various regions in gas parameter 
space. We have rejected from further consideration all solutions where the 
denser component is also required to be the hotter, as we consider this to
be physically unlikely on the scales observed. Although the various line
ratios are very similar, in particular the isotope ratios for M~83 
appear to be systematically somewhat lower than for NGC~6946. For this
reason, and also in order to demonstrate the possible variation in model  
parameters, we have listed characteristic solutions for both galaxies
separately. The quality of each solution can be judged by comparing the 
calculated model line ratios in Table 5 with those observed in Table 4.
The single-component fit is included for comparison. 
The densest and coolest component has a fairly well-determined density of 
3000 $\cc$ and an even better determined column density $N(CO)$/d$V$ = 
6-10 $\times 10^{16} \cm2$ irrespective of density. The low-density 
component ($10^{2}-10^{3} \cc$) must have higher {\it column} 
densities $N(CO)$/d$V$ = 1--10 $\times 10^{17} \cm2$ but its precise 
temperature is difficult to determine as long as the relative emission 
ratios of the components are a free parameter. 

The observed C$^{\circ}$ and C$^{+}$ intensities are modelled with the same
{\it radiative transfer} model. For both we assume the CO-derived, 
two-component solutions to be valid as far as kinetic temperatures, $\h2$ 
densities and filling factors are concerned. We may then solve for 
C$^{\circ}$ and C$^{+}$ column densities. The column density of the hotter 
component is usually well-determined, but that of the cooler component is 
more or less degenerate. Rather than a single solution, a range of possible 
column density solutions is found. These are constrained by requiring similar 
velocity dispersions (of about 3--5 km s$^{-1}$) for both hot and cold 
components and by requiring the resulting total carbon column densities to 
be consistent with the {\it chemical} model solutions presented by Van 
Dishoeck $\&$ Black (1988). These models show a strong dependence of the 
$N(C)/N(CO)$ column density ratio on total carbon and molecular hydrogen 
column densities.

The very strong [CII] intensity observed in M~83 exceeds that expected
from the CO-derived solutions. It thus implies the presence of ionized 
carbon in high-density molecular volumes poorly represented by CO emission. 
Consequently, in Model 6 (Table 5) we ascribed essentially all [CII] 
emission to a gas with density $n(\h2) = 10^{4} \cc$ at temperature 
$T_{\rm kin}$ = 60 K whereas in Model 7 we assumed $n(\h2) = 3000 \cc$ 
at $T_{\rm kin}$ = 150 K. 

In order to relate total carbon (i.e. C + CO) column densities to those of 
molecular hydrogen, we have estimated [C]/[H] gas-phase abundance ratios 
from [O]/[H] abundances. Both galaxies have virtually identical
central abundances [O]/[H] = 1.75 $\times$ 10$^{-3}$ Zaritsky et al. 1994;
Garnett et al. 1997). Using results given by Garnett et al. (1999), notably 
their Figs. 4 and 6, we then estimate carbon abundances [C]/[H] = 
1.45$\pm$0.5 $\times$ 10$^{-3}$. As a significant fraction of all carbon 
will be tied up in dust particles, and not be available in the gas-phase, 
we adopt a fractional correction factor $\delta_{\rm c}$ = 0.27 (see for
instance van Dishoeck $\&$ Black 1988), so
that $N_{\rm H}$ = [2$N(H_{2})$ + $N(HI)$] $\approx$ 2500 [$N(CO)$ + $N(C)$] 
with a factor of two uncertainty in the numerical factor. 

The results of our model calculations are given in Table 6, which presents
beam-averaged column densities for both CO and C (C$^{\rm o}$ and C$^{+}$)
and the $\h2$ column densities derived from these. 
Table 6 also lists the total mass estimated to be present in the central 
molecular concentration ($R < 300$ pc) obtained by scaling the $\h2$
column densities with the $J$=2--1 $L_{tot}/\int T_{\rm mb}$d$V$ ratio 
from Table 3, and the face-on mass densities implied by
hydrogen column density and the galaxy inclination. Beam-averaged 
{\it neutral} carbon to carbon monoxide column density ratios are 
$N(C^{\rm o})/N(CO) \approx 0.9\pm0.1$ for both NGC~6946 and M~83, 
somewhat higher than the values 0.2--0.5 found for M~82, NGC~253 and M~83 
(White et al. 1994; Israel et al. 1995; Stutzki et al. 1997; Petitpas $\&$ 
Wilson 1998).

\subsection{The center of NGC~6946}

Notwithstanding the significant differences between the model
parameters, the hydrogen column densities, masses and mass-densities 
derived in Table 6 are very similar. The [CI] and [CII] line and 
the far-infrared continuum (Smith $\&$ Harvey 1996) intensities 
suggest that they predominantly arise in a medium of density close to 
$10^{4} \cc$ subject to a radiation field log $G_{\rm o}$ = 1 -- 1.5 
(cf. Kaufman et al. 1999). Emission from the molecules CS, H$_{2}$CO 
and HCN has been detected from the CO peaks in Fig. 2 (Mauersberger 
et al. 1989; H\"uttemeister et al. 1997; Paglione et al. 1995; 1997); 
their intensities likewise indicate a density $n(\h2) \approx 10^{4} 
\cc$ which is only provided by models 2 and 3 which we consider to be 
preferable. Note that the single-component CO fit (model 4), which we
have already rejected, also does not fit the C$^{\rm o}$ and C$^{+}$ 
intensities predicted by the PDR models (Kaufman et al. 1999). The 
high-density component probably corresponds to the molecular cloud 
complexes that are the location of the present, mild starburst in the 
center of NGC~6946 (Telesco et al. 1993; Engelbracht et al. 1996). It 
represents about a third of the total molecular mass, and contributes 
a similar fraction to the observed $J$=2--1 CO emission. The low-density
component has a temperature in the range $T_{\rm kin}$ = 60 -- 150 K,
and a density of order $n(\h2) \approx \cc$. This is conformed by a
reanalysis of the midinfrared $\h2$ measurements by Valentijn et al. 
(1996). The $J$=2--0 S(0) $\h2$ line intensity at 28 $\mu$m is entirely
consistent with these values for an ortho/para ratio of two (P.P. van der 
Werf, private communication). However, in order to also match the observed 
$J$=3--1 S(1) line strength at 17 $\mu$m, a small amount of high-temperature 
molecular gas with $T_{\rm kin} \approx 500 K, n(\h2) \approx 5000 \cc$ need
be present as well, but with a mass no more than a few per cent of
the mass given in Table 6. Our measurements are insensitive to such a 
component.

We thus conclude that the total mass of molecular gas within $R$ = 0.5
kpc from the nucleus of NGC~6946 is 18$\pm$3 million solar masses; this 
is about 1.5 per cent of the dynamical mass, so that the total mass of
the inner part of the galaxy must be completely dominated by stars. 
No more than a quarter of all hydrogen is HI; most is in the form of $\h2$. 
Between 15 and 25$\%$ of all hydrogen is associated with ionized carbon 
and almost equal amounts with neutral carbon and CO. Madden et al. (1993) 
reach very similar conclusions from C$^{+}$ mapping of NGC~6946, but find 
different masses. Part of this difference arises in our use of two
components rather than a single component. Another important difference
between this and other studies is our use of the gas-phase carbon abundance
rather than an assumed conversion factor to obtain hydrogen column densities
and masses. For NGC~6946, this results in effective conversion factors of 
the order of $X$ = 1 $\times$ 10$^{19}$ $\h2$ mol $\cm2$/$\kkms$,
which is more than an order of magnitude lower than traditionally assumed
values. The difference greatly exceeds the uncertainty of a factor of two 
or three associated with the carbon abundance, illustrating the danger of 
using `standard' conversion factors in centers of galaxies where conditions 
may be very different (higher metallicities, higher temperatures) from those
in galaxy disks.

The observed CO temperatures are typically a factor of 15 {\it lower} than
the model brightness temperatures, implying that only a small fraction of
the observing beam is filled by emitting material. We find small beam-filling
factors for the molecular material of order 0.06 -- not very dependent on 
choice of model. However, velocity-integrated intensities are a factor
of two or three {\it higher} than that of a model cloud, implying that the
average line of sight through NGC~6946 contains two or three clouds at
various velocities.

\subsection{A GMC in the bulge of NGC~6946}

\begin{figure}
\unitlength1cm
\begin{minipage}[b]{3.94cm}
\resizebox{4.2cm}{!}{\includegraphics*{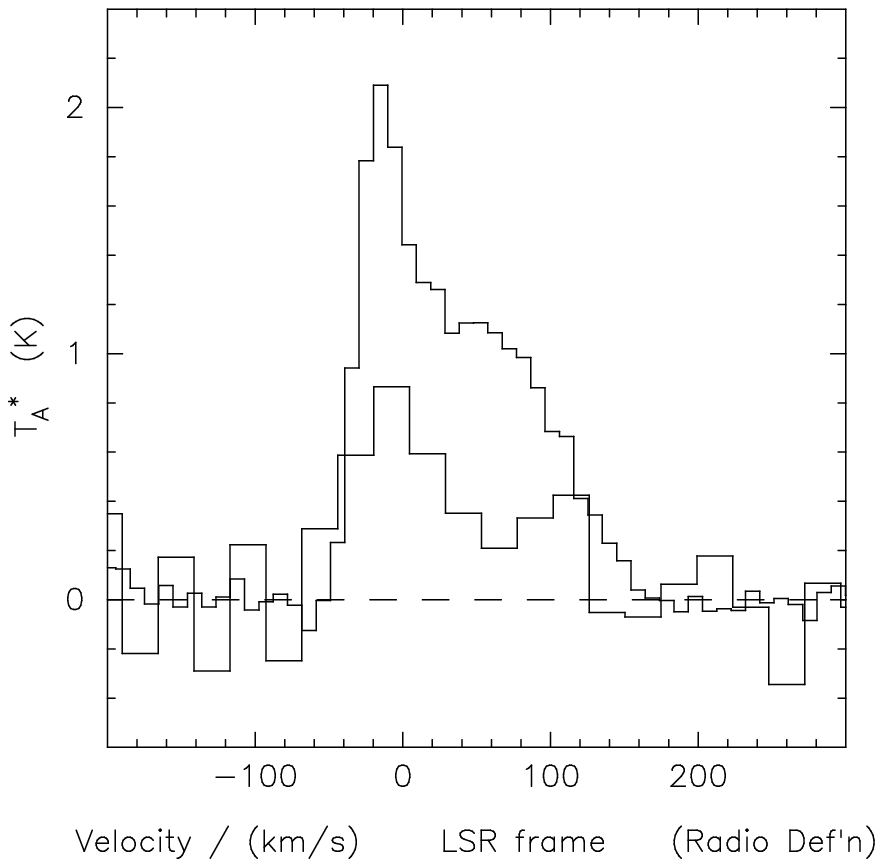}}
\end{minipage}
\hfill
\begin{minipage}[t]{3.94cm}
\resizebox{4.2cm}{!}{\includegraphics*{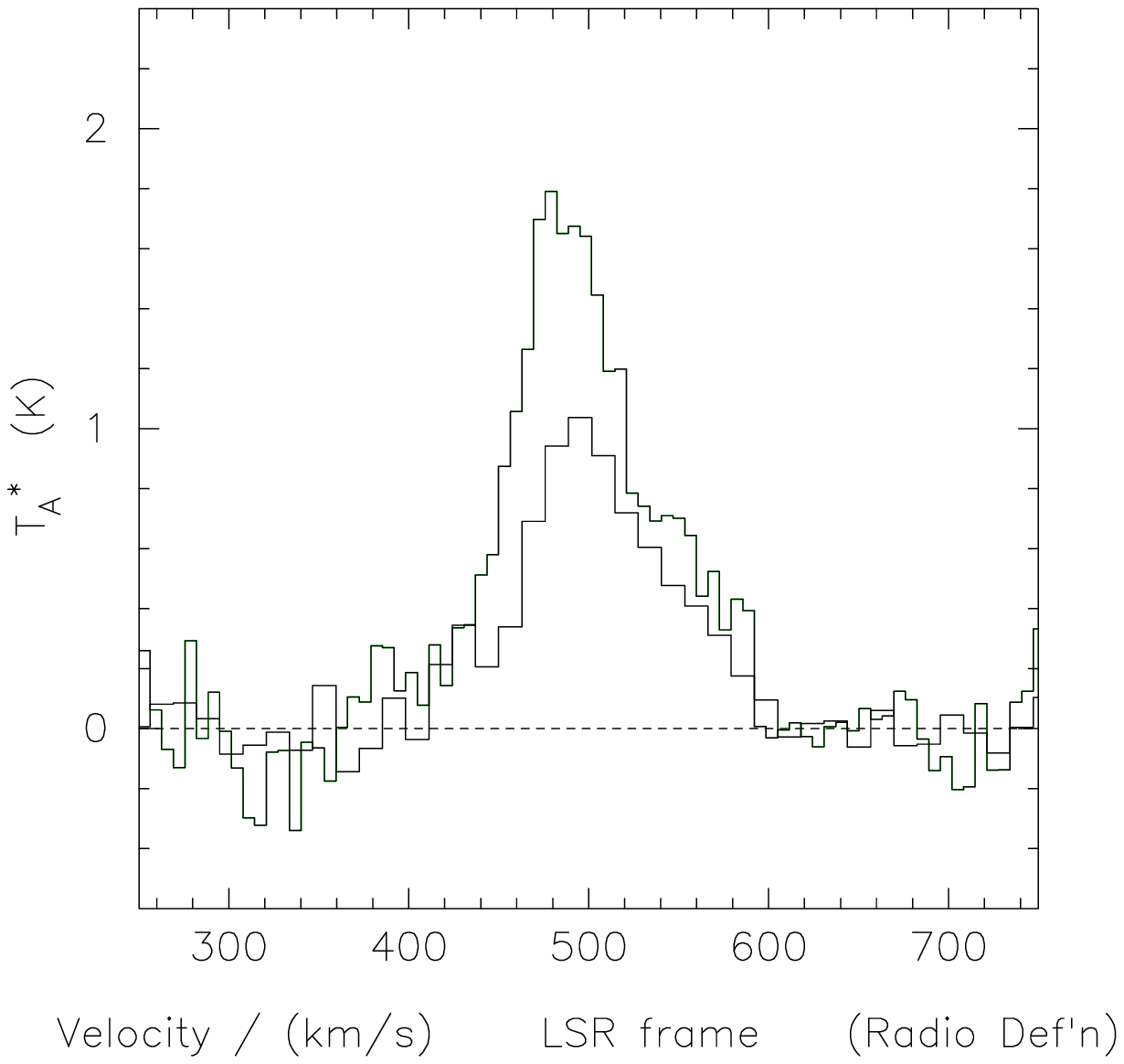}}
\end{minipage}
\caption[]
{$J$=2--1 emission profiles at offset positions; vertical scale is in 
$T_{\rm mb}$. Left: NGC~6946 at $\Delta \alpha, \Delta \delta$ = +10$''$, 
+10$''$. Right: M~83 at $\Delta \alpha, \Delta \delta$ = +7$''$, -7$''$. 
In both cases, the $\13co$ profile, multiplied by a factor of seven, is 
included as the lower of the two curves. 
}
\end{figure}

The eastern extension seen in our $J$=3--2 and $J$=2--1 maps is caused 
by a discrete cloud at $\Delta \alpha, \Delta \delta$ = +12$''$, +6$''$.
At this position, CO profiles show a strong, relatively narrow spike
asymmetrically superposed on the weaker broad profile from the
more extended emission (Fig. 6). This spike can also be discerned in
$J$=1--0 CO profiles published by Sofue et al. (1988) and in the $J$=3--2 
and $J$=4--3 CO profiles by Nieten et al. (1999). By subtracting the
broad emission, we have attempted to determine the parameters of this
cloud. We find a deconvolved size of about 400 pc along the major axis
and $\leq$ 160 pc (i.e. $\leq$ 260 pc deprojected) along the minor axis. 
Its deprojected distance to the nucleus is about 350 pc. Peak emission 
occurs at $V_{LSR}$ = -20 $\kms$, and the linewidth is $\Delta$V(FWHM) = 
30 km/s. These results suggest that the object is a molecular cloud complex 
in the bulge of NGC~6946, comparable to the Sgr B2 complex in the Milky Way.
Although the subtraction procedure is not accurate enough to obtain
good line ratios, these do not appear to be very different from those
of the major central concentration. They indicate a total mass $M(\h2)
\approx 3\times 10^{6}$ M$_{\odot}$ for the complex. Most of this mass
should be at a kinetic temperature of about 10 K, but about 15$\%$ of
the total mass should experience a temperature of order 100 K.

\subsection{The center of M~83}

Not surprisingly in view of the very similar CO line ratios, the
radiative transfer solutions for M~83 do not differ much from those
for NGC~6946 (Table 5). The major difference is found in Table 6,
and is caused by the much stronger [CII] emission. With model 5,
the [CII] intensity can be reproduced using the CO derived gas
parameters, but only if in the hot 100 K component essentially
all (94$\%$) carbon is in the ionized atomic form C$^{+}$; very 
little CO can be left. Use of the CO two-component parameters requires 
solutions with implausibly high C$^{+}$ column densities for models 6 
and 7. As already mentioned in the previous section, we have instead 
assumed that the [CII] emission from M~83 mostly samples conditions 
inbetween those of the two components, i.e. those at the interface of 
hot, tenuous and colder, denser gas. 

Whichever model is preferred, typically 50$\%$--65$\%$ of all carbon in
the center of M~83 must be in ionized form.  Because of this,
and the rather low HI column density observed towards the center of
M~83, molecular hydrogen column densities must be quite high, of
order 5--7 $\times 10^{21} \cm2$. Although only a relatively small
fraction of all $\h2$ is related to CO emission, the conversion factor
is nevertheless higher for M~83 than for NGC~6946: $X = 0.25 \times
10^{20}\cm2/\kkms$, but still well below the Galactic standard
value.

The models are consistent with densities $10^{3} \cc$ subject to 
radiation fields log $G_{\rm o}$ = 2 implied by comparing the CO, 
[CI] and [CII] line and far-infrared continuum (Smith $\&$ Harvey 1996) 
intensities with the PDR models given by Kaufman et al. (1999). Few 
density estimates from other molecules exist. Paglione et al. (1997) 
estimate $n(\h2) \leq 10^{3} \cc$ from HCN $J$=3--2 and $J$=1--0 
measurements, whereas the beam-corrected ratio $I(CO)/I(HCN)$ = 9 
($J$=1--0) from Israel (1982) suggests $n(\h2) \approx 
3\times 10^{4} \cc$ (see Mauersberger $\&$ Henkel 1993, their Fig. 4).

An important difference between NGC~6946 and M~83 is that the strong [CII]
emission characterizing the latter cannot be explained by assuming
that only relatively modest amounts of carbon monoxide have been 
photodissociated into atomic carbon. The considerably stronger starburst 
in M~83 (Gallais et al. 1991; Telesco et al. 1993; Turner $\&$ Ho 1994) 
has apparently created a PDR-zone in which large amounts of high-temperature, 
high-density ionized carbon gas have largely replaced efficiently eroded 
CO clouds, so that a significant fraction, of order 80$\%$, of the 
molecular hydrogen in this PDR-zone is effectively not sampled by CO 
emission. Dense, [CII] emitting gas is thereby a major contributor to
the total gas content of the center of M~83. The actual contribution
is somewhat uncertain because of the uncertainty in [CII] gas temperature.
If we take $T_{kin}$ = 250 K and $n(\h2) = 10^{4} \cc$ instead of the
actual values adopted, the resulting masses for models 6 and 7 in Table 
6 would be about 60$\%$ of the listed values.

We conclude that the total amount of molecular gas in the center of M~83 
(20$\pm$10 million solar masses) is very similar to that in NGC~6946
(18$\pm$3 million solar masses). As in the case of NGC~6946, this is of 
order 1--2 per cent of the dynamical mass, so that the mass of gas is 
negligible with respect to the stellar mass. About 6$\%$ all hydrogen 
is HI; the remainder must be in the form of $\h2$. About {\it half} of 
all hydrogen is associated with ionized carbon; the other half is mostly 
associated with CO. We thus confirm the predominant role for C$^{\rm o}$ 
that was already found by Crawford et al. (1985) and Stacey et al. (1991). 
As for NGC~6946, we note that the total molecular mass found in the central
region ($R < 0.5$ kpc) is much less than suggested by others on the
basis of assumed conversion factors.

In M~83, observed CO temperatures are typically a factor of 7.5 (Model 5)
to 15 (Model 7) lower than the model brightness temperatures, indicating 
beam-filling factors for the molecular material of order 0.12 -- 0.06, 
i.e. larger than for NGC~6946. At the same time, the velocity-integrated 
intensity is a factor of two to five larger than that of a model cloud, 
implying the presence of that number of clouds in an average line of sight 
through M~83. Although the central gas masses  in NGC~6946 and M~83 are 
very similar, the face-on mass density in the center of M~83 is more than 
double that of NGC~6946.

\section{Conclusions}

\begin{enumerate}

\item Maps of the central arcmin of the starburst galaxies NGC~6946
and NGC~5236 (M~83) in various transitions of $^{12}$CO and $^{13}$CO,
and in [CI] confirm the compact nature of the central molecular gas 
emission in both galaxies. Most of this gas is within a few hundred
parsec from the nucleus.  Major-axis position-velocity diagrams
show that in both galaxies the circumnuclear molecular gas is  in 
very rapid solid-body rotation. The steepness of the velocity gradient
only becomes apparent at the higher spatial resolutions.

\item Relative $^{12}$CO, $^{13}$CO and C$^{o}$ line intensities observed 
in matched beams are virtually identical in NGC~6946 and M~83, although
the [CII] line intensity in the literature is much stronger by a factor
of 7 in M~83. Spatially integrated line intensity ratios do not differ 
much from those obtained in the central 21$''$ beam, except for [CI] which 
is either more strongly concentrated towards the nucleus than CO or
insufficiently mapped.

\item The velocity-integrated $^{12}$CO intensities in both galaxies
decrease only slowly with increasing rotational level. The 
intensities in the $J$=1--0, $J$=2--1, $J$=3--2 and $J$=4--3 transitions 
are in the ratio of 1 : 1 : 0.65 : 0.45 respectively. Both galaxies
have observed $^{12}$CO/$^{13}$CO isotopic ratios of about 11, 9.5 and 
12.5 in the first three transitions.

\item The intensity of the neutral carbon line at 492 GHz relative to
$J$=2--1 (and $J$=1--0) $^{12}$CO is about 0.2 in both galaxy centers.
The relative intensity of the ionized carbon line is 0.08 for NGC~6946
and 0.55 for M~83.

\item The resemblance of the relative CO line intensities suggests
that the dense interstellar medium in both galaxies is very similar. 
However, the great difference in [CII] intensities shows that a
reliable picture is only obtained by observing and modelling both
atomic carbon and carbon monoxide lines. 

\item Modelling of the observed line ratios suggest a multi-component
molecular medium in both galaxies. In NGC~6946, a dense component
with $n(\h2) \approx 0.3-1.0 \times 10^{4} \cc$ and $T_{kin} \approx$
30 K is present together with a significantly less dense $n(\h2) 
\approx 0.5-1.0 \times 10^{3} \cc$ and hotter $T_{kin} \approx$ 
100--150 K component. Atomic carbon column densities appear to be
about 1.5 times the CO column density. The gas in M~83 may likewise
be approximated by two similar components.The denser is both somewhat 
less dense ($n(\h2) \approx 0.3 \times 10^{4} \cc$) and somewhat 
hotter ($T_{\rm kin}$ = 60 K) than its counterpart in NGC~6946. The 
more tenuous component is practically identical to its counterpart 
in NGC~6946. M~83 is more affected by CO dissociation, as its atomic 
carbon to CO ratio is about four. In both starburst centers, most 
of the molecular mass (about two thirds) is associated with the PDR
hot, relatively tenuous phase. In M~83, a significant molecular
gas volume must be associated with ionized carbon rather than CO.

\item With an estimated gas-phase [C]/[H] abundance of 4 $\times 
10^{-4}$, the centers of NGC~6946 and M~83 contain almost identical
total (atomic and molecular) gas masses of about 3 $\times
10^{7}$ $M_{\odot}$ within $R$ = 0.3 kpc. Peak face-on gas mass densities 
are, however, rather different: typically 45 $M_{\odot}\, pc^{-2}$ for 
NGC~6946 and almost three times higher, 115 $M_{\odot}\, pc^{-2}$ for 
M~83. The central molecular concentration in M~83 is denser and hotter 
than the one in NGC~6946.

\end{enumerate}
\acknowledgements

We are indebted to Ewine van Dishoeck and David Jansen for providing us 
with their detailed radiative transfer models and to Paul van der Werf
for his willingness to reanalyse the ISO $\h2$ measurements within the context
of our results. Fabienne Casoli kindly
supplied us with an IRAM $J$=2--1 $^{12}$CO map for comparison with
our data. We thank the JCMT personnel for their support and help in 
obtaining the observations discussed in this paper. 

\end{document}